\documentclass[%
 superscriptaddress,
 preprint,
 12pt,
 nofootinbib,
 amsmath,amssymb,
 aps,
]{revtex4-2}

\usepackage[utf8]{inputenc}
\DeclareUnicodeCharacter{2212}{-}
\usepackage{graphicx}
\usepackage{dcolumn}
\usepackage{bm}

\usepackage{bm}
\usepackage{braket}
\usepackage{tikz}
\usepackage{ulem}
\usepackage{hyperref}

\definecolor{lime}{HTML}{A6CE39}
\DeclareRobustCommand{\orcidicon}{
	\hspace{-3mm}
	\begin{tikzpicture}
	\draw[lime, fill=lime] (0,0) 
	circle [radius=0.16] 
	node[white] {{\fontfamily{qag}\selectfont \tiny ID}};
	\draw[white, fill=white] (-0.0625,0.095) 
	circle [radius=0.007];
	\end{tikzpicture}
	\hspace{-3mm}
}

\foreach \x in {A, ..., Z}{\expandafter\xdef\csname orcid\x\endcsname{\noexpand\href{https://orcid.org/\csname orcidauthor\x\endcsname}
			{\noexpand\orcidicon}}
}

\begin{document}

\preprint{KEK-TH-2666, OU-HET-1247}

\title{Anatomy of finite-volume effect on hadronic vacuum polarization contribution to muon \texorpdfstring{$g-2$}{TEXT}}

\author{Sakura~Itatani\orcidA{}}
\email{itasaku@post.kek.jp}
\affiliation{
The Graduate Institute for Advanced Studies (SOKENDAI), Tsukuba, Ibaraki 305-0801, Japan
}%


\author{Hidenori~Fukaya\orcidB{}}
\email{hfukaya@het.phys.sci.osaka-u.ac.jp}
\affiliation{
Department of Physics, Graduate School of Science, Osaka University, Toyonaka, Osaka 560-0043, Japan
}%

\author{Shoji~Hashimoto\orcidC{}}
\email{shoji.hashimoto@kek.jp}
\affiliation{
The Graduate Institute for Advanced Studies (SOKENDAI), Tsukuba, Ibaraki 305-0801, Japan
}%
\affiliation{%
Institute for Particle and Nuclear Studies, 
High Energy Accelerator Research Organization (KEK),
Tsukuba, Ibaraki 305-0801, Japan}

\date{\today}

\begin{abstract}
Low-energy spectrum relevant to the lattice calculation of hadronic vacuum polarization contribution to muon anomalous magnetic moment $a_\mu^{\mathrm{HVP,LO}}$ is dominantly given by two-pion states satisfying L\"uscher's finite-volume quantization condition.
Finite-volume effects from those states may exhibit power-law dependence on the volume, contrary to an exponential suppression as suggested by chiral effective theory.
Employing the finite-volume state decomposition of Euclidean correlators, we systematically investigate the volume dependence and identify the different volume scalings depending on the region. 
Using phenomenological inputs for $\pi\pi$ phase shift and time-like pion form factor, we obtain an estimate for the finite-volume effects on $a_\mu^{\mathrm{HVP,LO}}$, which is consistent with previous works.
Numerical results are given for the ``window'' observables of $a_\mu$.
\end{abstract}

\maketitle


\clearpage
\section{Introduction}
\label{intro}

The finite-volume effect is one of the most important sources of systematic error in the lattice QCD computation of hadronic vacuum polarization (HVP) contribution to muon anomalous magnetic moment $a_\mu=(g_\mu-2)/2$. 
In particular, the isovector component at the leading order, written as $a_\mu^{\mathrm{HVP,LO}}$, gives the largest contribution and thus potentially involves the most significant systematic errors.
For example, in the BMW calculation in 2020 \cite{Borsanyi:2020mff} the finite-volume effect at their reference lattice of size $L_{\mathrm{ref}}\simeq$ 6.3~fm is estimated as $19(3)\times 10^{-10}$, which is much larger than the error of the current experimental average, $\sim 2\times 10^{-10}$ \cite{Muong-2:2023cdq}.
Given the limitation of the lattice volume one can simulate with currently available computational resources, a reliable estimate of the finite-volume effect is of great relevance to the test of the Standard Model through $a_\mu$. 
In fact, an estimated finite-volume correction is added to the result to obtain the final result in \cite{Borsanyi:2020mff}, and its uncertainty is estimated to be approximately the same size as the experimental error.

Some theoretical estimates of the correction due to the finite spatial and temporal extent of the lattice are available from chiral perturbation theory (ChPT) \cite{Aubin:2015rzx,Bijnens:2017esv,Aubin:2020scy,Aubin:2022hgm} or from more general approaches \cite{Hansen:2019rbh,Hansen:2020whp}, and a correction that is suppressed at least as $\exp(-m_\pi L)$ is found in the leading order, for the pion mass $m_\pi$ and spatial extent of the lattice $L$.
There is, on the other hand, an argument that suggests a dependence of power law at least from the region of large Euclidean time separation of the current correlator that enters in the evaluation of HVP \cite{Hansen:2019rbh,Hansen:2020whp}. 
This is understood as follows.
In a finite spatial volume, the energy levels are quantized according to L\"uscher's condition \cite{Luscher:1990ux} and their level spacing scales as an inverse power of $L$. 
A sum of the contributions from these quantized states could naturally develop a power-law dependence like $1/L^\alpha$, especially from the lowest end of the spectrum. 
This effect could become most important for large Euclidean time separations where only a few lowest energy levels dominate, and is relevant for the estimate of $a_\mu^{\mathrm{HVP,LO}}$, as it receives a significant contribution from the time separation of the order of the inverse muon mass $1/m_\mu\sim$ 2~fm.  
The transition of such power-law dependence to the exponential one as expected from other analyses and its numerical impact are still to be understood.

In this work, we systematically study the effect of finite volume on $a_\mu^{\mathrm{HVP,LO}}$ using the framework to construct finite-volume Euclidean correlation functions from the phenomenological input. 
The two-pion states in a finite volume are identified using L\"uscher's formula \cite{Luscher:1990ux}, and their transition matrix element from the vacuum is given using the Lellouch-L\"uscher factor \cite{Lellouch:2000pv,Lin:2001ek} as explicitly formulated in \cite{Meyer:2011um}. 
In a finite box, the rotational symmetry is violated and different angular momentum states can mix.
In the present case of the angular momentum $J=1$ $\pi\pi$ states, the states of $J=3$ can mix, for example.
We neglect such effects, assuming that their effect on the estimate of the finite-volume effect is not significant.

An estimate of $a_\mu^\mathrm{HVP,LO}(L)$ at a finite volume $L$ is then obtained using the time-momentum representation \cite{Bernecker:2011gh} from the Euclidean correlation functions. 
Phenomenological inputs are needed for the two-pion scattering phase shift $\delta_1^1(k)$ (for isospin $I=1$ and angular momentum $J=1$) and for the time-like pion form factor $F_\pi(s)$. 
In our analysis, we use the commonly adopted Gounaris-Sakurai model \cite{Gounaris:1968mw}.
We also assume that inelastic state contributions, such as those from four pions and other higher-energy states, can be neglected when estimating the finite-volume effect. 
This framework has been applied in \cite{Lin:2001ek,Francis:2013fzp,Feng:2014gba,Giusti:2018mdh,Borsanyi:2020mff,Ce:2022kxy,FermilabLatticeHPQCD:2023jof} to estimate the finite-volume effect; our work extends the framework to include essentially all energy $\pi\pi$ eigenstates (see Section~\ref{sec:sum_over_states} for details), while most of the previous works treated only a small number of lowest states so that the volume analyzed was limited or other methods had to be combined in the analysis.
    
Contributions from different length scales are commonly studied using the ``window'' quantities \cite{RBC:2018dos}, as described in the community white paper \cite{Aoyama:2020ynm}. 
Three length scales are considered: short-distance window (SD), intermediate window (W), and long-distance window (LD). 
Their region of Euclidean time integral in the time-momentum representation is defined by $[0,0.4]$~fm, $[0.4,1.0]$~fm, and $[1.0,\infty]$~fm, respectively, allowing a transition region of width 0.15~fm implemented with a sigmoid function. 
(Details are discussed in Section~\ref{sec:smeared_time_windows}.) 
So far, the simulation results from various groups have been confirmed to agree in the intermediate window (see \cite{Boccaletti:2024guq}, for example, for a summary of the results from \cite{Lehner:2020crt,Wang:2022lkq,Aubin:2022hgm,Ce:2022kxy,FermilabLatticeHPQCD:2023jof,RBC:2023pvn}). 
The intermediate window is designed such that the discretization effect and the finite-volume effect are less important. 
Similar comparison of lattice data in the long-distance window is still missing (see the recent results \cite{Blum:2024drk,Bazavov:2024eou}), and the subject of this work is the finite-volume effect that is expected to become the most significant in this window.

The dominant contribution in the long-distance region arises from the two-pion states, for which the interaction vanishes in the low-energy limit.
The long-distance physics is therefore expected to be well described by a non-interacting pion states as the zeroth order approximation. 
The two-pion contribution can be written analytically in that limit (for large Euclidean time separation), and it provides a reference for extrapolation towards the infinite-volume limit. 
This analysis is described in Section~\ref{sec:non-interacting_correlator}.

For more realistic estimates of the finite-volume effect, the resonance enhancement of the $\gamma\pi^+\pi^-$ vertex due to the $\rho$ meson in the time-like pion form factor may play an important role. Such effect could not be fully incorporated within the framework of ChPT, and the method used in this work is needed.

We focus on the leading finite-volume effects on the isovector channel, while contributions from other flavours or from isospin breaking are ignored. 
Finite-volume effects for these states are expected to be much smaller as the mass of involved states is larger.

The structure of this paper is as follows. 
Section~\ref{sec:formulae} summarizes the standard formulae in the lattice calculation of $a_\mu^{\mathrm{HVP,LO}}$. 
Then, in Section~\ref{sec:non-interacting_correlator} we outline the analytic calculation applicable for non-interacting pions. 
The analysis is extended to the interacting case in Section~\ref{sec:interacting_correlator}, and the results for the finite-volume effects for $a_\mu^{\mathrm{HVP,LO}}$ including the predictions for the window quantities are presented in Section~\ref{sec:result}.
Conclusions with some discussion are given in Section~\ref{sec:conclusion}.
\footnote{Compared to the previous version posted on arXiv, we corrected an error in the calculation of the Lellouch-L\"uscher factor and fixed precision issues in the numerical computation of the zeta function.
This affected the main results and conclusions.}

\section{Two-pion contribution to \texorpdfstring{$a_\mu^{\mathrm{HVP,LO}}$}{TEXT}}
\label{sec:formulae}

In this section, we summarize the necessary formulae for the evaluation of $a_\mu^{\mathrm{HVP,LO}}$ to define the notations. 

\begin{figure}[tb]
    \centering
    \includegraphics[width=0.3\linewidth]{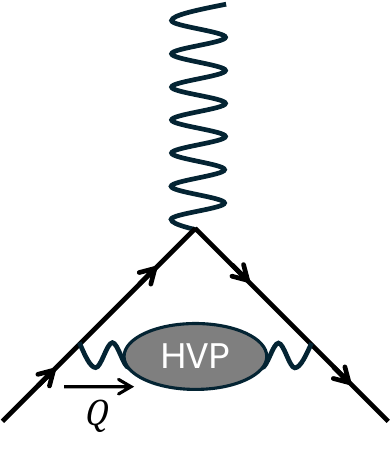}
    \caption{Leading-order hadronic vacuum polarization contribution to muon $g-2$}
    \label{fig:HVP}
\end{figure}

The leading-order (the order of $\alpha^2$ of fine structure constant $\alpha$) hadronic vacuum polarization contribution to muon anomalous magnetic moment, $a_\mu^{\mathrm{HVP,LO}}$, is written as an integral of the vacuum polarization function $\Pi(Q^2)$ over space-like momentum squared $Q^2$:
\begin{equation}
    a_\mu^{\mathrm{HVP,LO}}=4\alpha^2 \int_0^\infty\! dQ^2\, K_E(Q^2) [\Pi(Q^2)-\Pi(0)]
    \label{eq:momentum-rep}
\end{equation}
with a known kernel function $K(Q^2)$ \cite{Blum:2002ii} obtained from an integral of the vertex function in Figure~\ref{fig:HVP}.
It is given as
\begin{equation}
    K_E(Q^2)
    = \frac{1}{m^2_\mu} \hat{Q}^2 \cdot Z(\hat{Q}^2)^3  \frac{1-\hat{Q}^2 Z(\hat{Q}^2)}{1+\hat{Q}^2 Z(\hat{Q}^2)^2}
    \label{eq:KE}
\end{equation}
with
\begin{equation}
    Z(\hat{Q}^2) = -\frac{\hat{Q}^2-\sqrt{\hat{Q}^4 + 4\hat{Q}^2}}{2\hat{Q}^2},
    \quad \hat{Q}^2 \equiv \frac{Q^2}{m^2_\mu}
\end{equation}
for the muon mass $m_\mu$.
The space-like momentum kernel function $K_E(Q^2)$ behaves as $1/(\hat{Q}^2)^3$ for large $\hat{Q}^2$, so the contribution from the large energy region is highly suppressed.

The vacuum polarization function $\Pi(Q^2)$ is defined through
\begin{widetext}
\begin{equation}
    (Q_\mu Q_\nu-Q^2\delta_{\mu\nu})\Pi(Q^2) = 
    \int\!d^4x\, e^{iQx} 
    \bra{0} j_\mu^{\mathrm{em}}(x) j_\nu^{\mathrm{em}}(0) \ket{0}
\end{equation}
\end{widetext}
with electromagnetic current $j_\mu^{\mathrm{em}}(x)$.
By a Fourier transform in the time direction, the representation (\ref{eq:momentum-rep}) can be rewritten in the form of an integral over Euclidean time \cite{Bernecker:2011gh}, which is more convenient for lattice QCD calculations:
\begin{equation}    
    a_\mu^\mathrm{HVP,LO}
    = 4\alpha^2 m_\mu \int_0^{\tau_c} d\tau\,
    \tau^3 G(\tau) \tilde{K}_E(\tau).
    \label{eq:time-momentum}
\end{equation}
An Euclidean correlator $G(\tau)$ is also obtained by a Fourier transform as
\begin{equation}
    G(\tau) \equiv \int\! d\bm{x}\, \langle 0|j_z^{\mathrm{em}}(\tau,\bm{x}) j_z^{\mathrm{em}}(0) |0\rangle,
    \label{eq:euclidean_corr_def}
\end{equation}
where we take the electromagnetic currents $j_\mu^{\mathrm{em}}(\tau,\bm{x})$ in a spatial direction $\mu=z$. 
It is projected to zero spatial momentum, and separated in Euclidean time $\tau$. 
When obtained in a finite volume, we denote the correlator by $G(\tau,L)$, and the corresponding $a_\mu^{\mathrm{HVP,LO}}$ obtained through (\ref{eq:time-momentum}) as $a_\mu^{\mathrm{HVP,LO}}(L)$. 
The kernel function $\tilde{K}_E(\tau)$ is written in terms of the corresponding Euclidean momentum-space representation $K_E(\omega^2)$ (\ref{eq:KE}) as \cite{Blum:2002ii} 
\begin{widetext}
\begin{equation}
   \tilde{K}_{E}(\tau) \equiv \frac{2}{m_\mu \tau^3} \int_0^\infty \frac{d\omega}{\omega} K_{E}(\omega^2) \left[ \omega^2 \tau^2 - 4\sin^2 \frac{\omega \tau}{2} \right].
   \label{eq:kernel}
\end{equation}
\end{widetext}

\begin{figure}[tb]
   \centering
   \includegraphics[width=0.6\linewidth]{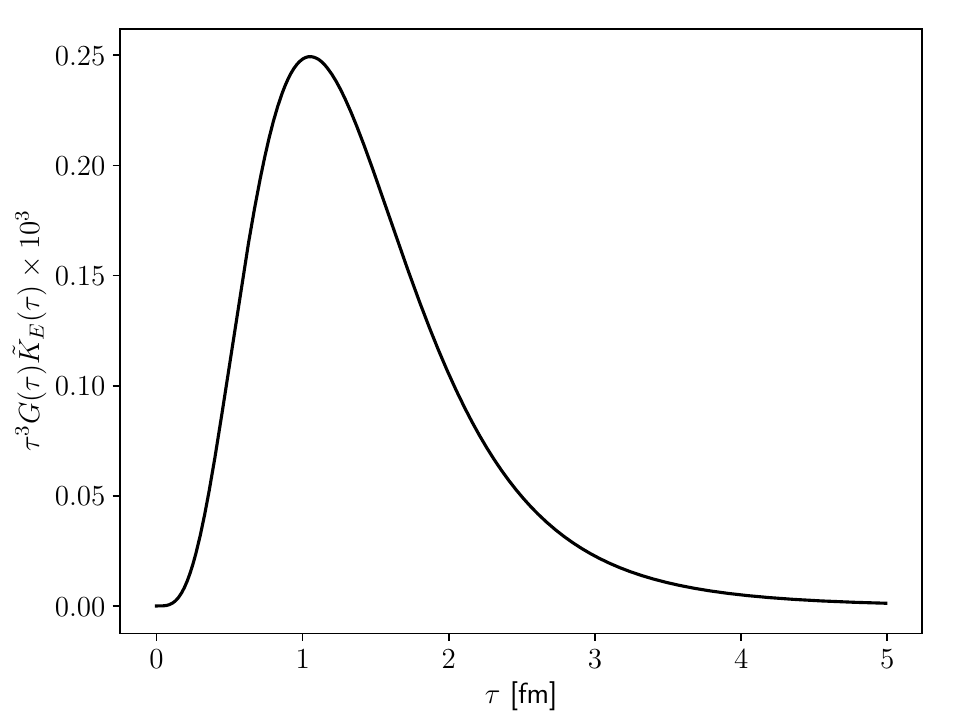}
    \caption{Integrand of the time momentum representation (\ref{eq:time-momentum}). The Gounaris-Sakurai model for the time-like pion form factor is used as described in the following sections.}
    \label{fig:integrand}
\end{figure}
    
The upper limit $\tau_c$ of the integral (\ref{eq:time-momentum}) should be set to infinity, but in practical lattice calculations the integral is cut off at a certain value of $\tau_c$.  
As shown in Figure~\ref{fig:integrand}, the integrand peaks around $\tau=$ 1~fm and rapidly decays towards larger $\tau$. 
The contributions from the time intervals [0,1]~fm, [1,2]~fm, [2,5]~fm are about 34\%, 39\%, 27\%, respectively, and that from the region beyond 5~fm is below 1\%.
The long-distance (LD) region beyond 1~fm (with a smearing of 0.15~fm as discussed above) \cite{Aoyama:2020ynm} roughly accounts for 60\% of the total $a_\mu^{\mathrm{HVP,LO}}$. 

In the long-distance region the Euclidean correlator is dominated by two-pion states and written as
\begin{equation}
    G(\tau,L) = \sum_n \left| \braket{0|j_z^{\mathrm{em}}|\pi\pi,n}_V \right|^2 e^{-E_{\pi\pi,n}\tau}
    \label{eq:correlator_decomposed}
\end{equation}
in a finite volume $V=L^3$. 
Here, the isovector channel is assumed and an isospin index for the two-pion state $\ket{\pi\pi,n}_V$ is suppressed.
The states are discretized in a finite volume, and the energy of $n$-th state $\ket{\pi\pi,n}_V$ is denoted as $E_{\pi\pi,n}$. 

In the infinite volume limit, where the two-pion states are simply labeled by the momenta $\bm{k}$ and $\bm{k}'$ of the pions in the asymptotic state, the matrix elements in (\ref{eq:correlator_decomposed}) are expressed in terms of the pion form factor $F_\pi(s)$:
\begin{widetext}
\begin{equation}
   \bra{0} j_\mu \ket{\pi_{\bm{k}}\pi_{\bm{k}'};\mathrm{in}} 
   = - \bra{\pi_{\bm{k}}\pi_{\bm{k}'};\mathrm{out}} j_\mu \ket{0} 
   = e^{i\delta_1^1(\bm{k}-\bm{k}')} (\bm{k}-\bm{k}')_\mu F_\pi(s).
   \label{eq:pipi-matrix_element}
\end{equation}
\end{widetext}
Here we consider the case of the center-of-mass frame $\bm{k}+\bm{k}'=0$. 
The momentum transfer is then $s=4(m_\pi^2+\bm{k}^2)$.
The phase shift of the scattering is denoted as $\delta_1^1(\bm{k}-\bm{k}')$.

We describe the method to construct the finite-volume correlation function $G(\tau,L)$ from phenomenological inputs for $F_\pi(s)$ and $\delta_1^1(\bm{k})$ in Section~\ref{sec:interacting_correlator}.

\section{Two-pion contribution in the non-interacting case}
\label{sec:non-interacting_correlator}

Before analyzing more realistic cases including interactions between pions, we consider the case of non-interacting pions, in order to gain some idea about possible asymptotic functional forms of the Euclidean correlator and $a_\mu^{\mathrm{HVP,LO}}(L)$.
It should provide a reasonable approximation at low energy, which is most relevant to the finite-volume effect because the pion interaction vanishes in the (massless and) low-energy limit.

\subsection{Current correlator}
For non-interacting pions, the two-pion states in (\ref{eq:correlator_decomposed}) are identified by their relative momentum $\bm{k}$.
The pion form factor in (\ref{eq:pipi-matrix_element}) is $|F_\pi(s)|=1$ and the phase shift is $\delta_1^1=0$. 
For a finite volume $V=L^3$, the Euclidean time correlation function is written as (see Appendix A.3 of \cite{Francis:2013fzp} as well as Section 2.2.3 of \cite{Hansen:2020whp})
\begin{equation}
   G(\tau,L)=\frac{1}{L^3} \sum_{\bm{k}}
   \frac{k_z^2e^{-2\sqrt{\bm{k}^2+m_\pi^2}\tau}}{\bm{k}^2+m_\pi^2},
   \label{eq:non-int_corr}
\end{equation}
where the sum is taken over $k_i=2\pi l_i/L$ with integers $l_i$.  
In order for a numerical evaluation, this expression is useful when $\tau/L \gg 1$ where the summation over $\bm{k}$ converges quickly. 
When $\tau/L\sim 1$ or smaller, the following expression is more adequate for numerical evaluation.

Using the Poisson resummation formula for an arbitrary function $f(p)$
\begin{widetext}
\begin{equation}
   \sum_l f\left(\frac{2\pi l}{L}\right) = 
   \int\! dp\, \sum_l \delta\left(p-\frac{2\pi l}{L}\right) f(p) = 
   L \sum_n\int\!\frac{dp}{2\pi}\, f(p)e^{inpL},
\end{equation}
\end{widetext}
we obtain
\begin{equation}
   G(\tau,L)=\sum_{\bm{n}} \int\!\frac{d^3p}{(2\pi)^3}\,\frac{p_z^2 e^{-2 \sqrt{\bm{p}^2+m_\pi^2}\tau}}{\bm{p}^2+m_\pi^2} e^{i\bm{n}\cdot \bm{p}L},
   \label{eq:GL}
\end{equation}
where three-component vector $\bm{n}=(n_x,n_y,n_z)$ consists of integers.
Invariance under $90^\circ$ rotations allows us to replace $p_z^2$ by $\bm{p}^2/3$, and using an integral over angular components of $\bm{p}$, which we denote by $\theta_p$ and $\varphi_p$,
\begin{widetext}
\begin{equation}
   \int_0^{2\pi} d\varphi_p \int _0^\pi d\theta_p \, \sin\theta_p\, e^{i |\bm{n}|pL \cos \theta_p} = \frac{4\pi\sin(|\bm{n}|p L)}{|\bm{n}|p L},
\end{equation}
\end{widetext}
we have
\begin{widetext}
\begin{equation}
   G(\tau,L) = \frac{m_\pi^3}{6\pi^2} \sum_{\bm{n}} \int_0^\infty dy \, \frac{y^4}{y^2+1} \frac{\sin(m_\pi L|\bm{n}|y)}{m_\pi L|\bm{n}|y} e^{-2 m_\pi \tau\sqrt{y^2+1}},
    \label{eq:non-int_corr_in-y}
\end{equation}
\end{widetext}
where we have changed a variable as $p(\equiv|\bm{p}|)=y m_\pi$.
Note that the $\bm{n}=\bm{0}$ part corresponds to the infinite-volume limit $L=\infty$.
The non-zero $|\bm{n}|$ terms correspond to the effect of pion wrapping around the finite-volume lattice.
In practice, the sum over $|\bm{n}|$ is truncated.

The effect of pion form factor, or the $\gamma\pi^+\pi^-$ vertex form factor, may be included by multiplying $|F_\pi(4(m_\pi^2+\bm{k}^2))|^2$ to the sum over $\bm{k}$ in (\ref{eq:non-int_corr}).
This amounts to inserting $|F_\pi(4m_\pi^2(1+y^2))|^2$ in (\ref{eq:non-int_corr_in-y}).
However, this may cause a problem when $|F_\pi(s)|^2$ is a rapidly changing function, as in the case for the $\rho$-meson resonance.
Since the integrand in (\ref{eq:non-int_corr_in-y}) contains a highly oscillating term $\sin(m_\pi L|\bm{n}|y)$, the integral could vary drastically for different $|\bm{n}|$  unless $|\bm{n}|$ is large enough to cancel out the oscillation.
The condition corresponds to $|\bm{n}|\gg 2\pi/\Delta k_\rho L$, where $\Delta k_\rho$ represents the momentum range in which the form factor varies rapidly.
For the physical $\rho$-meson resonance, $\Delta k_\rho$ is roughly $m_\rho\Gamma/8k_\rho\sim$ 40~MeV.
(Here, $\rho$ meson mass $m_\rho$ = 770~MeV and width $\Gamma$ = 150~MeV as well as the corresponding pion momentum $k_\rho=\sqrt{m_\rho^2/4-m_\pi^2}$ = 360~MeV are introduced.)
For example, when $L$ = 6~fm the condition implies $|\bm{n}|\gg 5$.
Thus, many terms have to be kept in the wrap-around expansion to introduce the resonance-like pion form factor.

Now we consider the finite-volume correction $\Delta G(\tau,L)\equiv G(\tau,L)-G(\tau,\infty)$:
\begin{widetext}
\begin{equation}
   \Delta G(\tau,L) = \frac{m_\pi^3}{12\pi^2} \sum_{\bm{n}\neq\bm{0}} \int_{-\infty}^\infty dy\, \frac{y^3}{y^2+1} \frac{e^{iym_\pi L|\bm{n}|} e^{-2m_\pi\tau\sqrt{y^2+1}}}{2i m_\pi L |\bm{n}|} \, + \text{c.c.}
\end{equation}
\end{widetext}
When $m_\pi\tau\gg 1$, the small $y^2$ region dominates the integral, and
\begin{widetext}
\begin{equation}
   \begin{split}
   \int_{-\infty}^\infty dy\, \frac{y^3}{y^2+1} & \frac{e^{iym_\pi L|\bm{n}|}}{2i m_\pi L |\bm{n}|} \exp\left[-2 m_\pi \tau\sqrt{y^2+1}\right] 
   \\
   \sim &  \int_{-\infty}^\infty dy\, \frac{y^3}{2i m_\pi L |\bm{n}|}\exp\left[-2 m_\pi \tau\left(1+y^2/2\right)+iym_\pi L|\bm{n}|\right]
   \\
   =& \frac{e^{-2 m_\pi\tau-\frac{m_\pi L^2 \bm{n}^2}{4\tau}}}{2i m_\pi L |\bm{n}|} \int_{-\infty}^\infty dy\,  y^3 \exp\left[-m_\pi \tau\left(y-\frac{iL|\bm{n}|}{2\tau}\right)^2 \right]
   \\
   =& \frac{e^{-2 m_\pi \tau-\frac{m_\pi L^2 \bm{n}^2}{4\tau}}}{2i m_\pi L |\bm{n}|}  \int_{-\infty}^\infty dz  \left(z + \frac{iL|\bm{n}|}{2\tau}\right)^3 \exp\left[-m_\pi\tau z^2 \right],
   \end{split}
\end{equation}
\end{widetext}
where the $y$ integral is analytically continued to the complex plane and replaced by that of $z\equiv y-\frac{iL|\bm{n}|}{2\tau}$.
Dropping odd functions in $z$, we obtain
\begin{widetext}
\begin{align}
    \Delta G(\tau,L) &=\frac{2 m_\pi^3}{12\pi^2}\sum_{\bm{n}\neq\bm{0}} \frac{e^{-2 m_\pi\tau-\frac{m_\pi L^2 \bm{n}^2}{4\tau}}}{4 m_\pi\tau}
    \int_{-\infty}^\infty dz  \left(3z^2 - \frac{L^2|\bm{n}|^2}{4\tau^2}\right)\exp\left[-m_\pi\tau z^2 \right]
    \nonumber\\
    &=\frac{m_\pi^3}{6\pi^2}\sum_{\bm{n}\neq\bm{0}} \frac{e^{-2 m_\pi \tau-\frac{m_\pi L^2 \bm{n}^2}{4\tau}}}{4 m_\pi \tau}
    \left(-\frac{3}{\tau}\frac{\partial}{\partial m_\pi} - \frac{L^2|\bm{n}|^2}{4\tau^2}\right) \int_{-\infty}^\infty dz  \exp\left[-m_\pi\tau z^2 \right]
    \nonumber\\
    &= \frac{m_\pi^3}{6\pi^2}\sum_{\bm{n}\neq\bm{0}} \frac{e^{-2 m_\pi\tau-\frac{m_\pi L^2 \bm{n}^2}{4\tau}}}{4 m_\pi \tau}
     \left(-\frac{3}{\tau}\frac{\partial}{\partial m_\pi} - \frac{L^2|\bm{n}|^2}{4\tau^2}\right)\sqrt{\frac{\pi}{m_\pi\tau}}
    \nonumber\\
    &= \sqrt{\frac{m_\pi}{\pi^3\tau^5}}\frac{e^{-2 m_\pi\tau}}{48} \sum_{\bm{n}\neq\bm{0}}
    \left(3 - \frac{m_\pi L^2|\bm{n}|^2}{2\tau}\right)\exp\left(-\frac{m_\pi L^2 \bm{n}^2}{4\tau}\right),
    \label{eq:finite-V_correction_upto_n}
\end{align}
\end{widetext}
which is Eq.~(A15) of \cite{Francis:2013fzp}.
Here, using
\begin{widetext}
\begin{equation}
    \sum_{\bm{n}\neq\bm{0}} \left(3 - \frac{m_\pi L^2|\bm{n}|^2}{2\tau}\right)\exp\left(-\frac{m_\pi L^2 \bm{n}^2}{4\tau}\right)
    = \left(3+2m_\pi\frac{\partial}{\partial m_\pi}\right) \sum_{\bm{n}\neq\bm{0}} \exp\left(-\frac{m_\pi L^2 \bm{n}^2}{4\tau}\right)
    \nonumber
\end{equation}
\end{widetext}
and a formula of the elliptic theta function
\begin{equation}
    \vartheta_3(0, q) = 2 \sum_{n=1}^\infty q^{n^2}+1 = \sum_{n=-\infty}^\infty q^{n^2},
    \nonumber
\end{equation}
we obtain a compact expression:
\begin{widetext}
\begin{equation}
    \Delta G(\tau,L) = \sqrt{\frac{m_\pi}{\pi^3\tau^5}}\frac{e^{-2 m_\pi\tau}}{48} \left(3+2m_\pi\frac{\partial}{\partial m_\pi}\right)
    \left[\left(
    \vartheta_3(0, e^{-\frac{m_\pi L^2}{4\tau}}) \right)^3-1\right].
    \label{eq:finite-V_correction_for_corr}
\end{equation}
\end{widetext}
For small $\zeta\equiv e^{-m_\pi L^2/4\tau}$ the elliptic theta function scales as $\vartheta_3(0,\zeta)\to 1+2\zeta+...$, so that $\Delta G(\tau,L)$ approaches zero as $\exp(-m_\pi L^2/4\tau)$. 
When $L\sim 4\tau$, this is consistent with the well-known estimate of the finite-volume scaling as $\exp(-m_\pi L)$. 
However, we find differences in the shorter and longer distances.
In the region of large time separation, $\tau\gtrsim L/4$, the exponential suppression becomes weaker than $\exp(-m_\pi L)$ due to a factor of $L/4\tau$, while for $\tau\lesssim L/4$ the suppression is stronger. Thus, the effect on $a_\mu^{\mathrm{HVP,LO}}$ depends on which region of $\tau$ one probes by the window quantity, for instance.

When $\zeta$ is not so small, say $\zeta>0.05$, the elliptic theta function is well approximated by $\vartheta_3(0,\zeta)\sim\sqrt{\pi/(-\ln\zeta)}$ \cite{Theta}.
It means that the square brackets in (\ref{eq:finite-V_correction_for_corr}) contain a term $(4\pi\tau/m_\pi)^{3/2} (1/L)^3$ for relatively small volumes $m_\pi L^2/4\tau \lesssim 3$.
It is remarkable that a power-law dependence appears in the large-volume scaling: $\sim 1/L^3$ for a fixed $\tau$. In (\ref{eq:finite-V_correction_for_corr}) this power-law dependence is accidentally canceled due to the term $(3+2m_\pi \partial/\partial m_\pi)$, but it can remain in general. In fact, we find a dependence like $1/(m_\pi L)^3$ for the interacting case (see  Fig.~\ref{fig:delta_a_mu_v.s._1/(m_piL)^3}, \ref{fig:delta_a_mu_v.s._1/(m_piL)^3_intvsnoint} in Sec.~\ref{sec:result}).

In general, the Euclidean correlator $\Delta G(\tau,L)$ has a complex dependence on volume and time separation due to the infinite sum over $\bm{n}$ in (\ref{eq:finite-V_correction_upto_n}).
The scaling of $a_\mu^{\mathrm{HVP,LO}}$ could, therefore, be complicated, as it is given by a weighted integral of $G(\tau,L)$ over $\tau$.
Namely, it may behave as a power law at relatively small $L$, say $m_\pi L\sim 6$, turn to exponential for larger $L$, and then decrease even faster for asymptotically large $L$.
The explicit calculation is shown below.

\subsection{Contribution \texorpdfstring{$a_\mu^{\mathrm{HVP,LO}}$}{amu}}
Let us define the finite-volume correction for $a_\mu^{\mathrm{HVP,LO}}$ as 
\begin{equation}
   \Delta a_\mu^{\mathrm{HVP,LO}}(L) \equiv a_\mu^{\mathrm{HVP,LO}}(L)-a_\mu^{\mathrm{HVP,LO}}(\infty),
\end{equation}
where $a_\mu^{\mathrm{HVP,LO}}(L)$ is obtained from the finite-volume Euclidean correlator $G(\tau,L)$ using (\ref{eq:time-momentum}).

\begin{figure}[tb]
   \centering
   \includegraphics[width=0.9\linewidth]{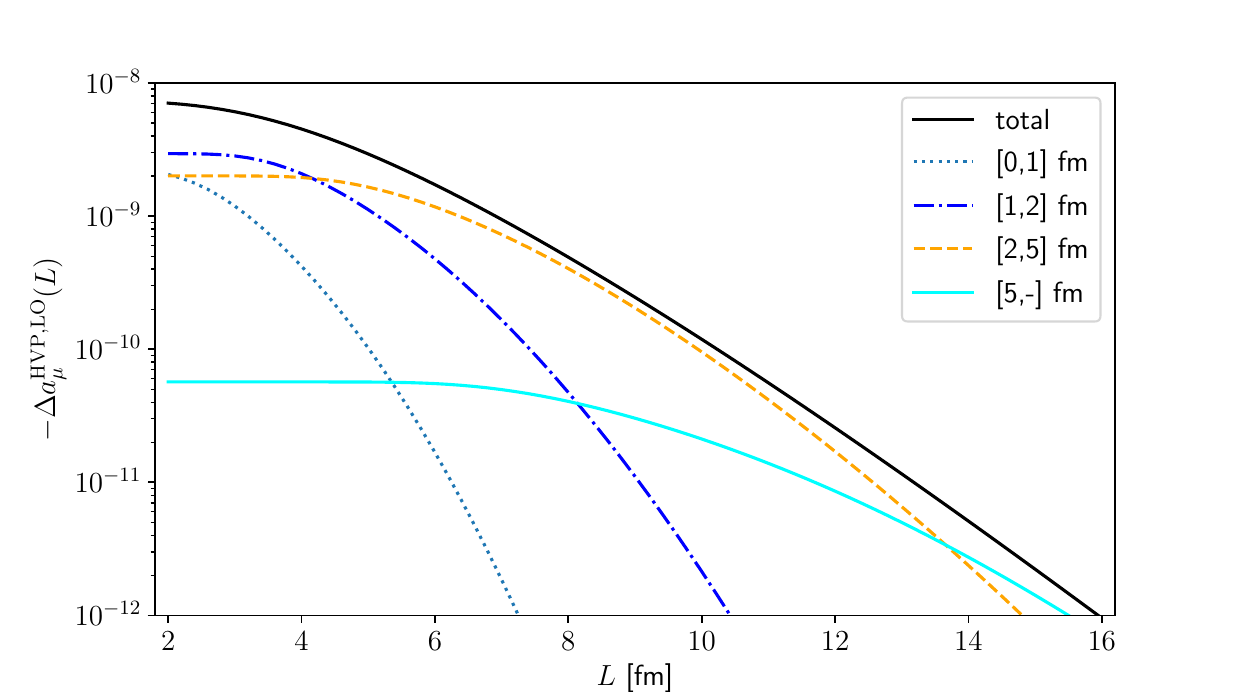}
   \caption{$|\Delta a_\mu^{\mathrm{HVP,LO}}(L)|=-\Delta a_\mu^{\mathrm{HVP,LO}}(L)$ constructed from non-interacting pion correlator (\ref{eq:finite-V_correction_for_corr}). The total (black curve) is divided into different Euclidean time contributions: [0,1]~fm (dotted), [1,2]~fm (dot-dashed), [2,5]~fm (dashed) and [5,$\infty$]~fm (cyan).  Pion mass $m_\pi$ is set to 140~MeV.}
   \label{fig:damu_log}
\end{figure}
    
Figure~\ref{fig:damu_log} shows $-\Delta a_\mu^{\mathrm{HVP,LO}}(L)$ constructed from the non-interacting two-pion correlator (\ref{eq:finite-V_correction_for_corr}), which is obtained under assumption $m_\pi\tau\gg 1$. 
Since this condition is not always satisfied in the integral (\ref{eq:time-momentum}), we also show a decomposition of the integral into different regions of $\tau$.
The different lines in Figure~\ref{fig:damu_log} represent contributions from a given Euclidean time range: [0,1]~fm (dotted), [1,2]~fm (dot-dashed), [2,5]~fm (dashed) and [5,$\infty$]~fm (cyan). 
The condition $m_\pi\tau\gg 1$ would be satisfied reasonably well for $\tau\simeq$ 2~fm or larger.
The finite-volume effect is negative and we plot $-\Delta a_\mu^{\mathrm{HVP,LO}}(L)$.

The most significant contribution comes from the Euclidean time range [1,2]~fm or [2,5]~fm depending on the volume size $L$. 
Namely, the range [1,2]~fm is dominant for $L\lesssim$ 4~fm, while the other range [2,5]~fm becomes far more important beyond $L\sim$ 5~fm.
Note that $a_\mu^{\mathrm{HVP,LO}}$ itself is dominated by the time interval $\sim 1/m_\mu$, which is typically around 1--2~fm.

We also find that $-\Delta a_\mu^{\mathrm{HVP,LO}}(L)$ decreases faster than exponential for large $L$.
The curve for each time range shows a curvature, most prominently for the smallest time range, [0,1]~fm, but for others as well. 
This is understood from the asymptotic form of (\ref{eq:finite-V_correction_for_corr}), {\it i.e.} $\exp(-m_\pi L^2/4\tau)$ with $\tau$ in the limited range. 
Since the integral (\ref{eq:time-momentum}) is dominated by the time range around $1/m_\mu$, one might think that the finite-volume effect scales as $\exp(-m_\pi m_\mu L^2/4)$ for asymptotically large $L$, but the explicit calculation shows a more complicated convolution of different time separations as demonstrated in the plot.

For the smaller $L$ region, on the other hand, the decrease of $|\Delta a_\mu^{\mathrm{HVP,LO}}(L)|$ towards larger volumes is much flatter, suggesting non-exponential or power-like behavior. 
This is again consistent with the expectation from the functional form of the correlator, as discussed above.

\begin{figure}[tb]
   \centering
   \includegraphics[width=0.9\linewidth]{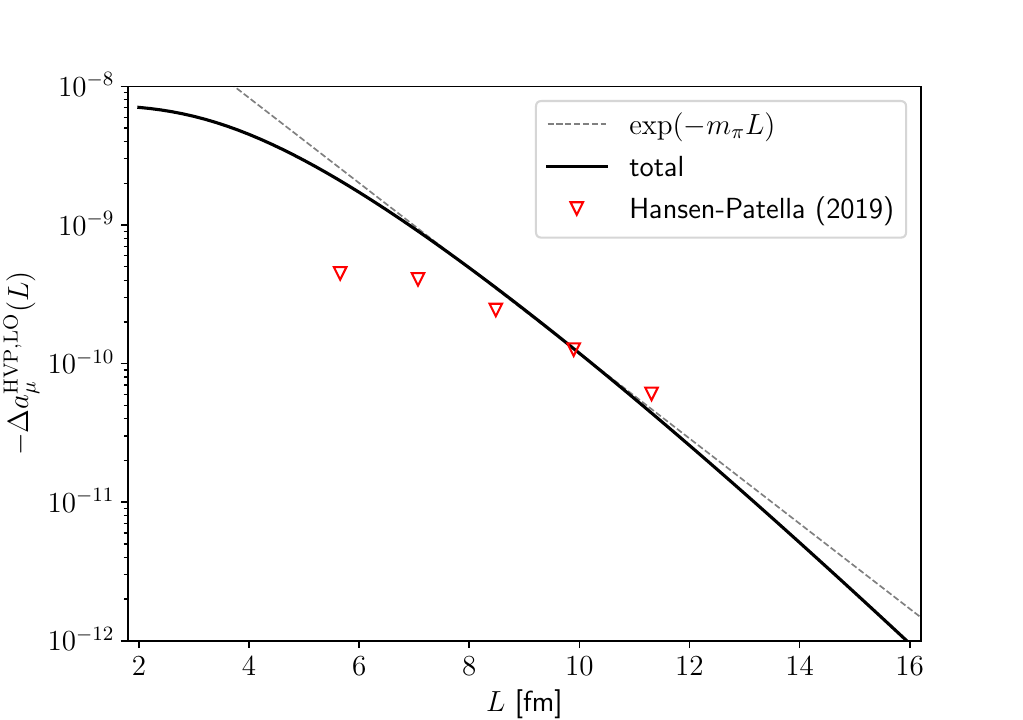}
   \caption{$-\Delta a_\mu^{\mathrm{HVP,LO}}(L)$ constructed from non-interacting pion correlator (\ref{eq:finite-V_correction_for_corr}). An estimate by Hansen and Petella 
   \cite{Hansen:2019rbh} (with an input $F_\pi(s)=1$) is shown by down-triangles. Also plotted is a line to show the slope of $\exp(-m_\pi L)$ (dashed line).}
   \label{fig:damu_log_withHP}
\end{figure}
 
In Figure~\ref{fig:damu_log_withHP} the total $-\Delta a_\mu^{\mathrm{HVP,LO}}(L)$ is compared with the previous result by Hansen and Petella \cite{Hansen:2019rbh} (with $F_\pi(s)=1$).
Our estimate is close to Hansen-Patella for $L\simeq$ 8--10~fm, but decreases more rapidly for large $L$ and increases more significantly for smaller volumes.
Also shown in Figure~\ref{fig:damu_log_withHP} is a line showing the slope of $\exp(-m_\pi L)$ as suggested in \cite{Hansen:2019rbh}. 
The magnitude of this exponential line is tuned such that it runs through the point near $L=$ 10~fm.
The estimate based on non-interacting pions shows a significantly different dependence on $L$ from the simple exponential suggested in previous studies.

\begin{figure}[tb]
   \centering
   \includegraphics[width=0.9\linewidth]{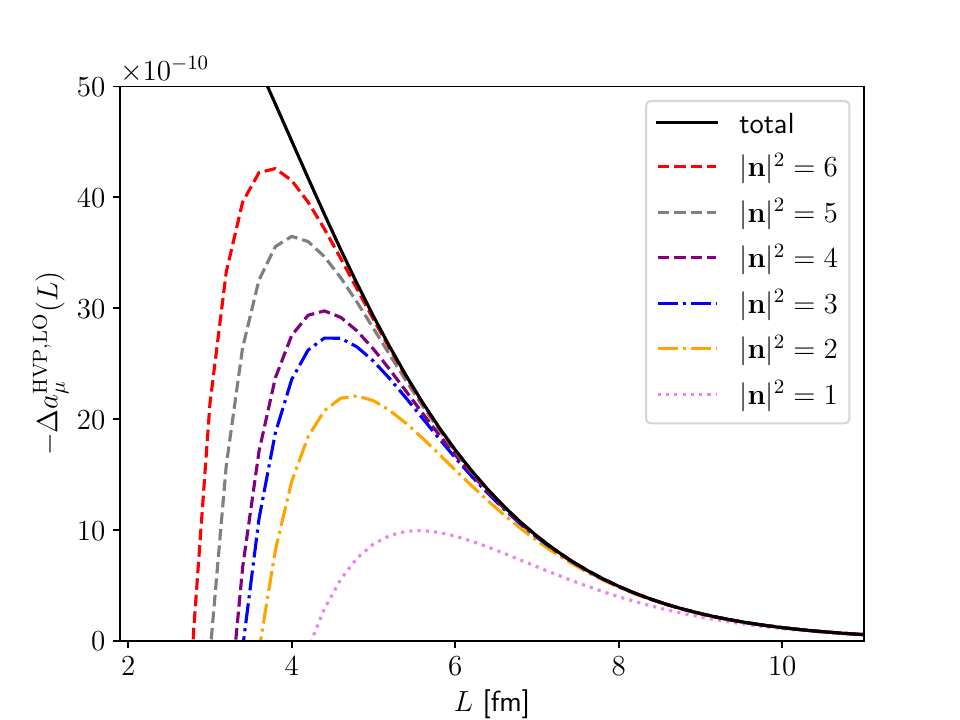}
   \caption{$-\Delta a_\mu^{\mathrm{HVP,LO}}(L)$ constructed from non-interacting pion correlator (\ref{eq:finite-V_correction_for_corr}). The contributions of wrap-around effects are added only up to a certain value of $|\bm{n}|^2$.}
   \label{fig:damu_decom}
\end{figure}

In chiral perturbation theory, the finite-volume effect is often estimated with a truncated sum over the integers $\bm{n}$ in (\ref{eq:GL}) that represent how many times pions wrap around the volume in each direction.
For each $|\bm{n}|$, the $L$-dependence is given by $e^{-m_\pi|\bm{n}|L}/\sqrt{m_\pi|\bm{n}|L}$ \cite{Aubin:2015rzx,Aubin:2020scy}, and the total is obtained by a sum over $|\bm{n}|$ taking into account a multiplicity for each $|\bm{n}|$.
Figure~\ref{fig:damu_decom} demonstrates the effect of the truncation.
We again use the approximation $m_\pi\tau\gg 1$, and the two-point function is given by (\ref{eq:finite-V_correction_for_corr}).
It turns out that the truncation at $|\bm{n}|=1$ leads to a significant underestimate even at $L\sim$ 6~fm or 8~fm.
The higher-order effects are also numerically important for small $L$'s.
This plot is obtained with the approximation $m_\pi\tau\gg 1$, but similar results would be expected for the full estimate.

\begin{figure}[tb]
   \centering
   \includegraphics[width=0.9\linewidth]{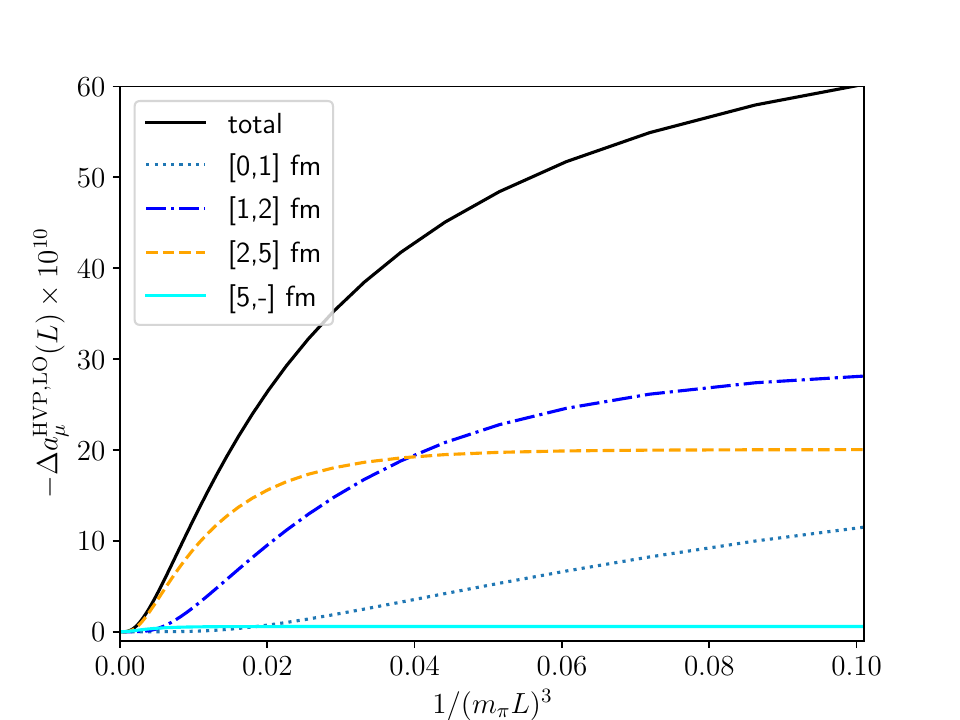}
   \caption{$-\Delta a_\mu^{\mathrm{HVP,LO}}(L)$ constructed from non-interacting pion correlator (\ref{eq:finite-V_correction_for_corr}) as a function of $1/(m_\pi L)^3$. Different lines are represented as in Figure~\ref{fig:damu_log}.}
   \label{fig:damu_L3}
\end{figure}
    
The same quantity $-\Delta a_\mu^{\mathrm{HVP,LO}}(L)$, without the truncation in $|\bm{n}|$, is plotted as a function of $1/(m_\pi L)^3$ in Figure~\ref{fig:damu_L3}. 
Near the infinite volume limit, the finite-volume correction vanishes exponentially or even faster as described above, while it increases significantly for finite volumes from $1/(m_\pi L)^3\sim$ 0.005, which corresponds to $m_\pi L\simeq$ 6. A linear increase can be seen for the contribution from [2,5]~fm, for example, below $1/(m_\pi L)^3\simeq 0.01$. Then, it saturates to a constant due to the cancellation mentioned above.

The overall size, of the order of $20\times 10^{-10}$ at the nominal volume size $m_\pi L\simeq$ 4, or $1/(m_\pi L)^3\simeq$ 0.016, is very significant compared to the target precision of about a few times $10^{-10}$.

\section{Two-pion contribution in the interacting case}
\label{sec:interacting_correlator}

In this section, we extend the estimate of the finite-volume effects to the case of interacting pions.
Instead of performing lattice QCD simulations, which is computationally expensive, we use the phenomenologically known $\pi\pi$ phase shift $\delta_1^1(\bm{k})$ and time-like pion form factor $F_\pi(s)$ to construct the Euclidean correlator.
In doing so, we neglect the inelastic effects due to four-pion states and higher, since their contribution to $a_\mu^{\mathrm{HVP,LO}}$ is sub-leading (10\% or less compared to the $\pi\pi$ contribution) and so is the finite-volume effect.
The Euclidean correlator (\ref{eq:correlator_decomposed}) on a finite volume is constructed from the energy spectrum $E_{\pi\pi,n}$ and the matrix element $\bra{0}j_z^{\mathrm{em}}\ket{\pi\pi,n}_V$. They are obtained by applying L\"uscher's condition for finite-volume states.

\begin{figure}[tb]
\centering
\includegraphics[width=0.9\linewidth]{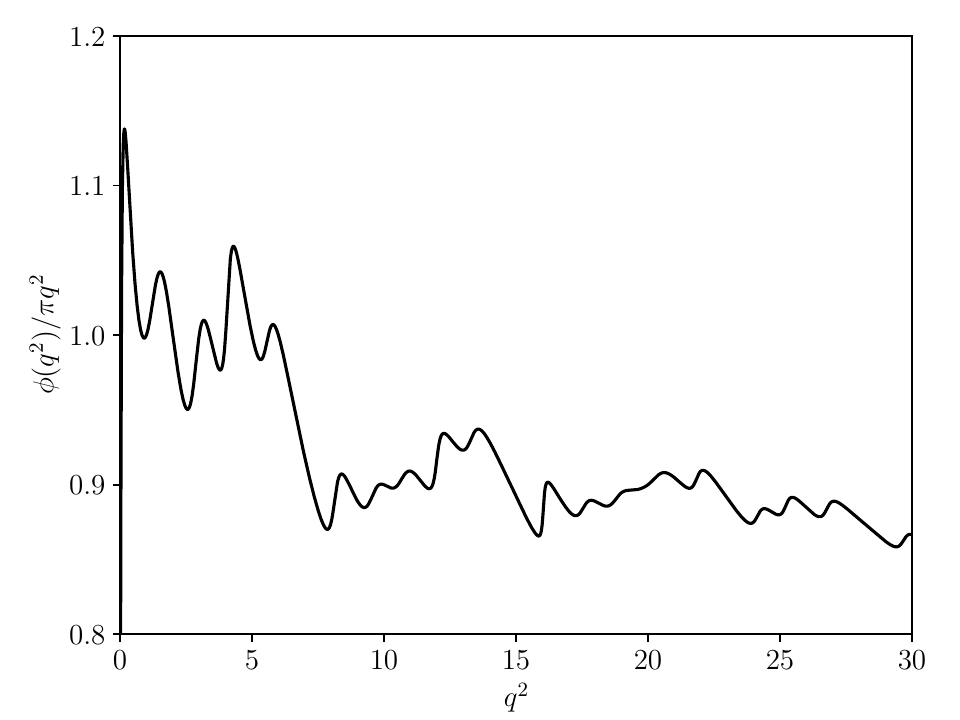}
\caption{$\phi(q)/\pi q^2$ computed for $q^2$ up to 35.0.
}
\label{fig:phi}
\end{figure}

\subsection{Finite-volume energy spectrum}
Two-pion spectrum on a finite volume satisfies Lüscher's condition \cite{Luscher:1990ux}:
\begin{equation}
   \delta(k) + \phi\left( \frac{kL}{2\pi} \right) = n\pi,
   \label{eq:Luscher_condition}
\end{equation}
where $k\equiv|\bm{k}|$ and $\delta(k)$ is a short-hand notation of $\delta_1^1(\bm{k})$ for the present case of $I=J=1$.
The two-pion state energy to satisfy (\ref{eq:Luscher_condition}) is obtained as $E_{\pi\pi,n}=2\sqrt{m_\pi^2+k^2}$ from $k$ for each $n$.
The function $\phi(q)$ is defined by $\tan\phi(q)=-\pi^{3/2}q/{\cal Z}_{00}(1;q)$ with Lüscher's generalized zeta function ${\cal Z}_{00}(s;q)=(1/\sqrt{4\pi})\sum_{\bm{l}}(\bm{l}^2-q^2)^{-s}$. 
On the right-hand side of (\ref{eq:Luscher_condition}), $n$ is a non-negative integer.

We compute $\phi(q)$ numerically up to $q=\sqrt{30.0}\approx 5.5$, beyond which we face a numerical instability.
We follow the prescription given in Appendix~C of \cite{Luscher:1990ux}.
The numerical integral necessary to compute $\mathcal{Z}_{00}(s;q)$ is performed with the best possible precision available in SciPy, and is cross-checked using Mathematica.
The numerical instability is observed as a strong cancellation among different terms when $q$ is too large.
The results are confirmed with the numerical table in \cite{Luscher:1991cf} for $q^2=$ 0.1--9.0.
Near $q^2=0$, especially below 0.1, our calculation of $\phi(q)$ agrees with (A.4) and (A.5) of \cite{Luscher:1991cf}.
The results for $\phi(q)/\pi q^2$ are shown in Figure~\ref{fig:phi}.
It is close to 1 at the first approximation, but a complicated structure is visible. 
In the following analysis, $n$ is limited to $n\leq 26$, which means that the highest energy $\pi\pi$ state for a given volume is limited.
We also assume that the contribution of higher partial waves of $l\geq 3$ can be ignored. 

\begin{figure}[tb]
    \centering
    \includegraphics[width=0.9\linewidth]{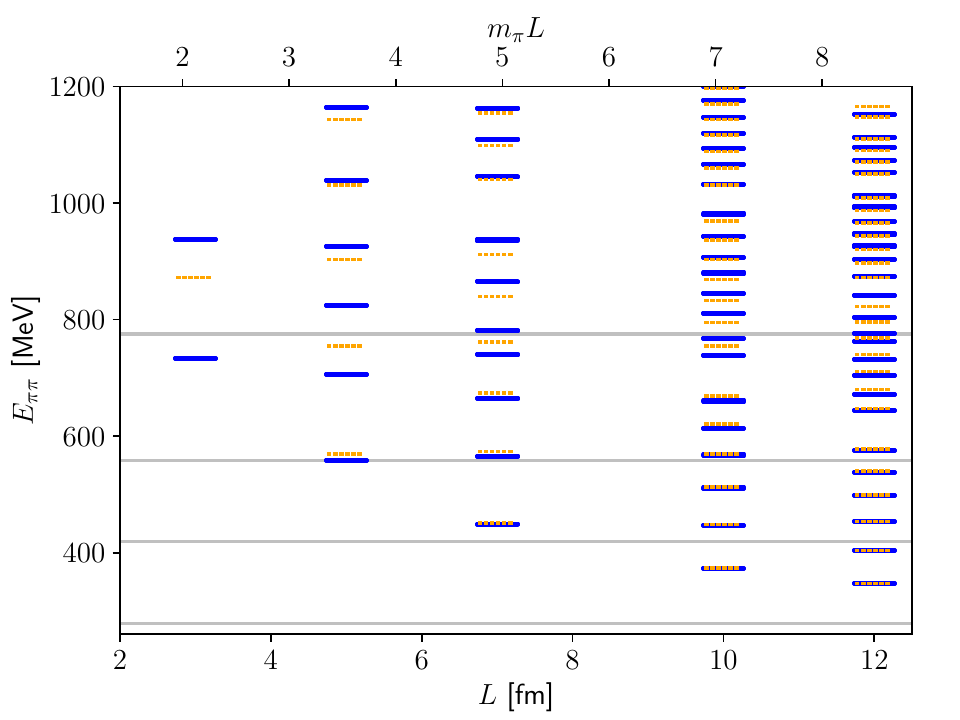}
    \caption{Energy spectrum of the $\pi\pi$ states on finite volumes of size $L$ = 3, 5, 7, 10 and 12~fm.
    The spectrum with (blue lines) and without (orange dotted) $\pi\pi$ interaction is shown for each volume.
    Gray lines correspond to $2m_\pi$, $3m_\pi$, $4m_\pi$, and $m_\rho$, from the bottom to top.
    The pion and rho meson masses are set to their physical values: $m_\pi$ = 139.6~MeV, $m_\rho$ = 775~MeV. 
    }
  \label{fig:energy_spectra}
\end{figure}

Figure~\ref{fig:energy_spectra} shows the energy spectrum of the $\pi\pi$ states on finite volumes. 
The $\pi\pi$ phase shift $\delta(k)$ for those of interacting spectrum (blue lines) is taken from Gounaris-Sakurai model (see below).
Non-interacting $\pi\pi$ spectrum is also shown (orange dotted lines), which is obtained by setting $\delta(k)=0$.
The highest energy state possible with $n\leq$ 26 is slightly below 1,200~MeV at $L$ = 12~fm, while the covered energy range extends much higher for smaller volumes.

\subsection{Matrix element from Gounaris-Sakurai model}
The matrix element $\bra{0}j_z^\mathrm{em}\ket{\pi\pi,n}_V$ is reconstructed from the elecromagnetic pion form factor $|F_\pi(s)|$ with Lellouch-Lüscher formula \cite{Lellouch:2000pv,Lin:2001ek} as described in \cite{Meyer:2011um}:
\begin{widetext}
\begin{equation}
   \left| F_\pi(E^2_{\pi\pi,n}) \right|^2
   = \left( q\phi'(q) + k\frac{\partial\delta(k)}{\partial k} \right) \frac{3\pi E^2_{\pi\pi,n}}{2k^5} \left| \braket{0|j_z^{\mathrm{em}}|\pi\pi,n}_V \right|^2.
   \label{eq:Lellouch-Luscher_formula}
\end{equation}
\end{widetext}
We assume the Gounaris-Sakurai (GS) model \cite{Gounaris:1968mw} for the pion form factor.
It is written as
\begin{equation}
   F^{\mathrm{GS}}_\pi(s)
   = \frac{- m^2_\rho - \Pi_\rho(0)}{s - m^2_\rho - \Pi_\rho(s)},
   \label{eq:GS_form_factor}
\end{equation}
where $\Pi_\rho(s)$ represents the $\rho$ meson self-energy as defined in the model.
In the GS model, $\Pi_\rho(s)$ is obtained using the (twice-subtracted) dispersion relation with an input 
\begin{equation}
   \mathrm{Im}\Pi_\rho(s) = -\frac{g_{\rho\pi\pi}^2}{6\pi} \frac{k^3}{\sqrt{s}},
\end{equation}
where $k=\sqrt{s/4-m_\pi^2}$.
The $\rho\pi\pi$ coupling $g_{\rho\pi\pi}$ is set using the experimentally available $\rho\to\pi\pi$ decay width as $\Gamma_{\rho\pi\pi} = (g_{\rho\pi\pi}^2/6\pi)(k_\rho^3/m_\rho^2)$.
In this analysis, we took $\Gamma_{\rho\pi\pi}$ = 149~MeV, which corresponds to $g_{\rho\pi\pi}= 5.976$. 

\begin{figure}[tb]
   \centering
   \includegraphics[width=0.9\linewidth]{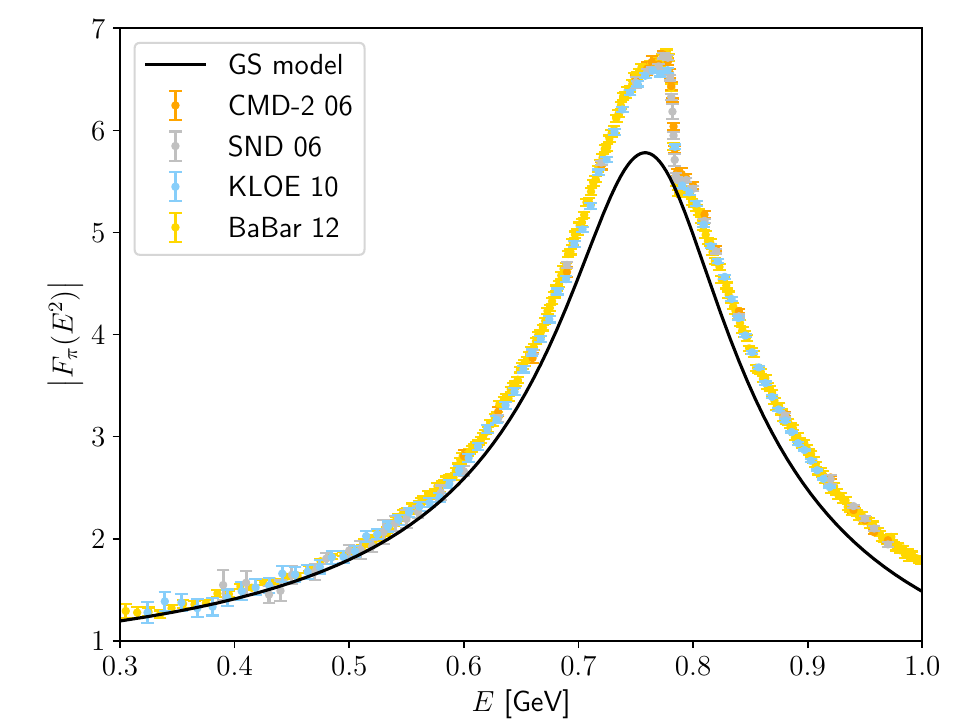}
   \caption{Time-like pion form factor from the experimental data of $e^+e^-\to\pi^+\pi^-$ (CMD-2 06 \cite{Aulchenko:2006dxz, CMD-2:2006gxt}, SND 06 \cite{Achasov:2006vp}, KLOE \cite{KLOE:2010qei}, BaBar \cite{BaBar:2012bdw}) and from Gounaris-Sakurai (GS) model (solid line).
   }
   \label{fig:formfactor_exp_vs_GS}
\end{figure}
    
The pion form factor (\ref{eq:Lellouch-Luscher_formula}) has a peak at the physical $\rho$-meson mass as shown in Figure~\ref{fig:formfactor_exp_vs_GS}.
The sudden drop near the peak due the $\rho$-$\omega$ mixing is not captured by the GS model (black curve), which contains the $I=1$ states only, but the overall shape of the experimental data is well reproduced.
The deficit of the GS model (black curve) compared to the experimental data is mainly attributed to the inelastic states \cite{Colangelo:2018mtw,Colangelo:2022prz}.
Since $|F_\pi(s)|$ is greatly enhanced due to the $\rho$ resonance, the contribution to $a_\mu^{\mathrm{HVP,LO}}$ can be larger compared to the non-interacting case $|F_\pi(s)|=1$.

\begin{figure}[tb]
   \centering
   \includegraphics[width=0.9\linewidth]{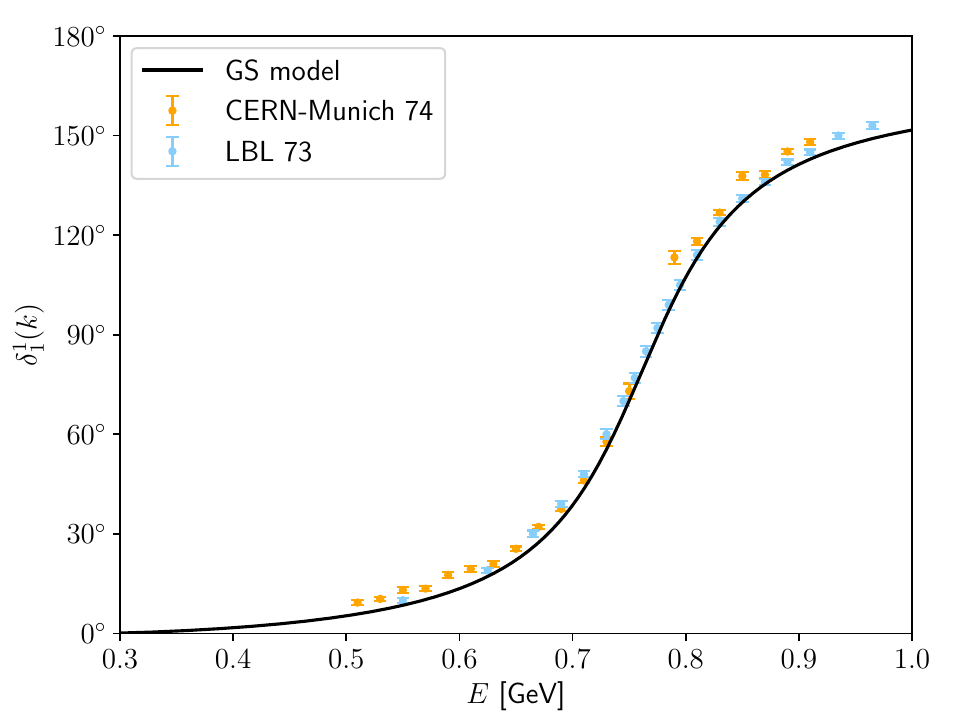}
   \caption{Phase shift from the experimental data (CERN-Munich 74 \cite{Estabrooks:1974vu} and LBL 73 \cite{Protopopescu:1973sh}), compared with Gounaris-Sakurai (GS) model (solid line).
   }
   \label{fig:phaseshift_exp_vs_GS}
\end{figure}

The phase shift $\delta(k)$ in the GS model is written as
\begin{equation}
    \frac{k^3}{\sqrt{s}}\cot\delta(k) = k^2 h(\sqrt{s})-k_\rho^2 h(m_\rho)+b(k^2-k_\rho^2),
    \label{eq:GS_phase_shift}
\end{equation}
where $k_\rho=\sqrt{m_\rho^2/4-m_\pi^2}$.
The function $h(\sqrt{s})$ and a constant $b$ are written as
\begin{align}
    h(\sqrt{s})
    & = \frac{2}{\pi} \frac{k}{\sqrt{s}} \ln \left( \frac{\sqrt{s}+2k}{2m_\pi} \right),
    \\
    b & = -h(m_\rho) - \frac{24\pi}{g^2_{\rho\pi\pi}} - \frac{2k^2_\rho}{m_\rho} \left. \frac{dh}{d\sqrt{s}} \right|_{\sqrt{s}=m_\rho},
\end{align}
and the $\rho$ meson mass is set with the condition
\begin{equation}
\left. \mathrm{Re}\Pi_\rho(s) \right|_{s = m^2_\rho} = 0,
\quad
\left. \frac{d}{ds} \mathrm{Re}\Pi_\rho(s) \right|_{s = m^2_\rho} = 0.
\end{equation}
Some details of the Gounaris-Sakurai model are found in the Appendix~A of \cite{Feng:2014gba}.

Figure~\ref{fig:phaseshift_exp_vs_GS} shows the experimental data for $\delta(k)=\delta_1^1(k)$ and the corresponding GS model.
The phase shift passes through $90^\circ$ around the mass of the $\rho$ meson.
The GS model describes the experimental data quite closely.
    
\subsection{Sum over the low-lying \texorpdfstring{$\pi\pi$}{TEXT} states}
\label{sec:sum_over_states}
We reconstruct the finite-volume Euclidean correlator $G(\tau,L)$ using (\ref{eq:correlator_decomposed}) with finite-volume energy $E_{\pi\pi,n}$ and matrix element $\braket{0|j_z^{\mathrm{em}}|\pi\pi,n}_V$ as described above.
Since the major contribution to $a_\mu^{\mathrm{HVP,LO}}(L)$ arises from energy levels that are not too high compared to $m_\mu$ due to the suppression of the kernel function $K_E(Q^2)$ in large $Q^2$, we expect that the sum over $n$ saturates rapidly beyond $E_{\pi\pi}\sim$ 1~GeV.
The finite-volume effect $\Delta a_\mu^{\mathrm{HVP,LO}}(L)$ should be even more dominated by low-energy states.

\begin{figure}[p]
   \centering
   \includegraphics[width=0.52\linewidth]{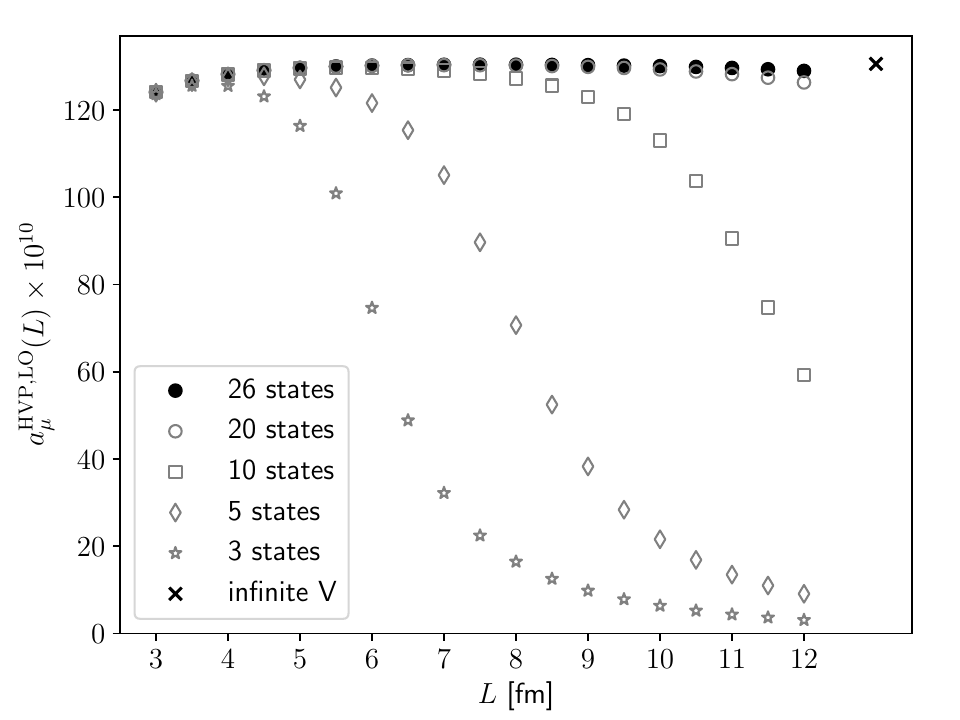}
   \includegraphics[width=0.52\linewidth]{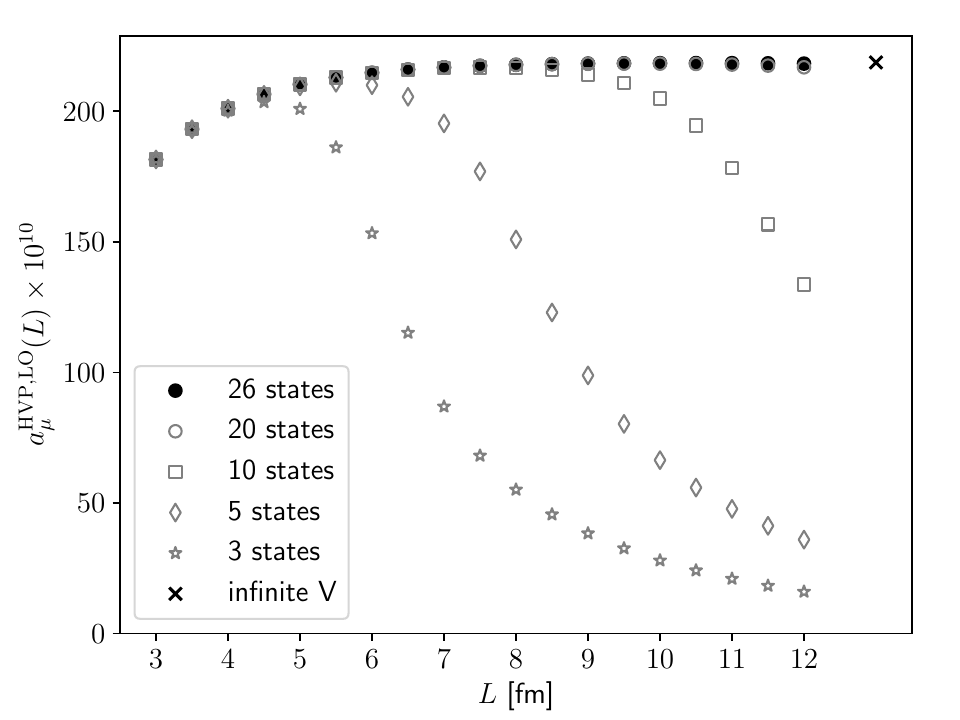}
   \includegraphics[width=0.52\linewidth]{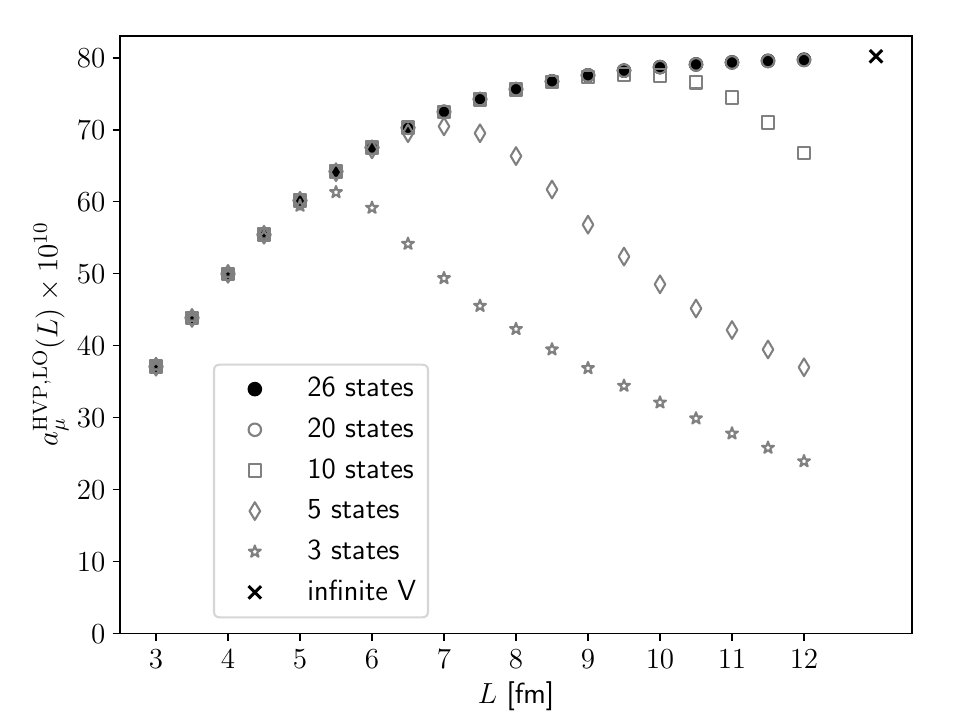}
   \caption{Saturation of the sum over $\pi\pi$ states for $a_\mu^{\mathrm{HVP,LO}}(L)$. Lowest $n$ states are summed to construct the correlator $G(\tau,L)$: $n$ = 3, 5, 10, 20, and 26. Results for partial time integrals in the range [0,1]~fm (top panel), [1,2]~fm (middle), [2,10]~fm (bottom) are shown as a function of $L$. Crosses represent the infinite volume limit including all possible states.}
   \label{fig:saturation}
\end{figure}

In Figure~\ref{fig:saturation} we plot $a_\mu^{\mathrm{HVP,LO}}(L)$ constructed from a finite number of low-lying $\pi\pi$ states.
That is, the sum (\ref{eq:correlator_decomposed}) is truncated at an upper limit $n$ = 3, 5, 10, 20, and 26 when constructing the finite-volume correlator $G(\tau,L)$.
The $\tau$ integral is divided into [0,1]~fm (top panel), [1,2]~fm (middle) and [2,10]~fm (bottom) as in the analysis of the non-interacting case.
The estimate of $a_\mu^{\mathrm{HVP,LO}}(L)$ is well saturated by a smaller number of states, even 3 or 5, for smaller volumes, $L\sim$ 3--4~fm, as expected from the corresponding energy spectrum (Figure~\ref{fig:energy_spectra}), {\it i.e.} the energy of the fifth lowest state is already above 1~GeV for $L$ = 3~fm and 5~fm.
Going to larger volumes, the saturation becomes slower, but we find that the difference between $n=20$ and 26 is very small: the symbols (open and filled circles) are nearly overlapping in the plot.
Numerically, the difference between $n=20$ and 26 is $3\times 10^{-10}$, $1\times 10^{-10}$, $0.1\times 10^{-10}$ for the time ranges [0,1]~fm, [1,2]~fm and [2,10]~fm, respectively, for the largest volume $L=$ 12~fm.

\begin{figure}[tb]
   \centering
   \includegraphics[width=0.9\linewidth]{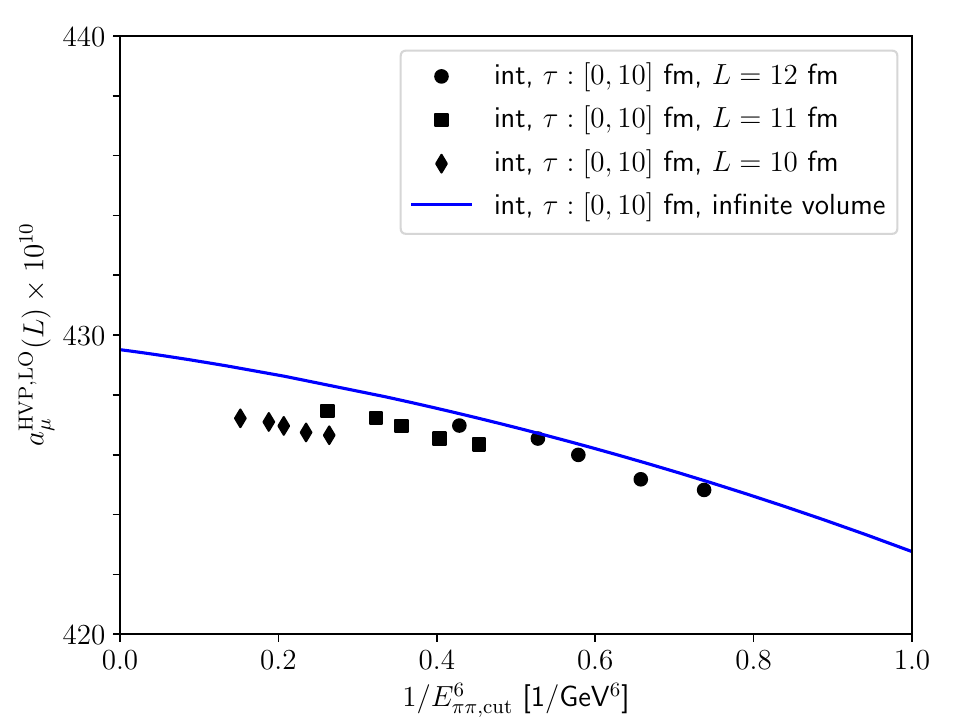}
   \caption{$a_\mu^{\mathrm{HVP,LO}}(L)$ reconstructed with a truncated sum over states in (\ref{eq:correlator_decomposed}). Results for $L$ = 12~fm (circles), 11~fm (squares) and 10~fm (diamonds) with $\tau_c$ = 10~fm are plotted as a function of $1/E_{\pi\pi,\mathrm{cut}}^6$, where $E_{\pi\pi,\mathrm{cut}}$ is the highest energy $E_{\pi\pi,n}$ included. 
   Data points correspond to $n$ = 22--26 (from right to left). 
   The infinite volume limit is shown by a blue curve.
   }
   \label{fig:extrapolation}
\end{figure}

The dependence on the upper limit of the energy sum, $E_{\pi\pi,\mathrm{cut}}$, can be understood from the dispersion integral representation of $a_\mu^{\mathrm{HVP,LO}}$, which is written as $\sim\int_0^\infty ds/s\, K(s) R(s)$ with the kernel $K(s)$ and the $R$ ratio $R(s)$ (see, {\it e.g.} \cite{Aoyama:2020ynm}). For large $s=E^2$, the kernel scales as $K(s)/s\sim 1/s^2$ and the two-pion contribution to $R(s)$ decreases as $1/s^2$, so we expect the cutoff dependence of $1/E_{\pi\pi,\mathrm{cut}}^6$. It should also be nearly independent of the volume, as it is a UV effect. Fig.~\ref{fig:extrapolation} shows $a_\mu^{\mathrm{HVP,LO}}(L)$ as a function of $1/E_{\pi\pi,\mathrm{cut}}^6$. The blue curve represents the dependence in the infinite volume limit, while the filled symbols show the truncated sum for $L=$ 10, 11, and 12~fm. We can confirm that the shape is nearly independent of the volume.

We therefore extrapolate the results for $a_\mu^{\mathrm{HVP,LO}}(L)$ towards $n\to\infty$ assuming the $E_{\pi\pi,\mathrm{cut}}$-dependence obtained in the infinite volume limit. In practice, we fit the 3--5 data points of the highest $E_{\pi\pi,\mathrm{cut}}$ with this scaling form (taking $E_{\pi\pi,\mathrm{cut}}=E_{\pi\pi,n}$) plus a constant shift representing the finite-volume effect. The deviation among the fits with 3--5 points is $0.07\times 10^{-10}$ for the worst case of $L=$ 12~fm and much smaller for 11~fm and 10~fm. We take an average of the three fit results as an estimate for $a_\mu^{\mathrm{HVP,LO}}(L)$. The lowest cutoff energy involved in the extrapolation is about 1.1~GeV for $L=$ 12~fm. We use the maximum deviation from the average as an estimate of systematic error. This estimate of error is added for both directions, positive and negative, to be conservative.

The infinite volume limit to compare (crosses in Figure~\ref{fig:saturation}) is obtained using the Euclidean correlator directly reconstructed from the pion form factor :
\begin{equation}
   G(\tau) = \int_0^\infty dE\, E^2 \rho(E^2) e^{-E\tau},
   \label{eq:correlator_infiniteV}
\end{equation}
where
\begin{equation}
   \rho(E^2) = \frac{1}{48\pi^2} \left( 1 - \frac{4m^2_\pi}{E^2} \right)^{3/2} \left| F^\mathrm{GS}_\pi(E) \right|^2
\end{equation}
is the spectral function representing the $\pi\pi$ states.

\subsection{Results for \texorpdfstring{$a_\mu^{\mathrm{HVP,LO}}$}{amu}}

\begin{figure}[tb]
   \centering
   \includegraphics[width=0.9\linewidth]{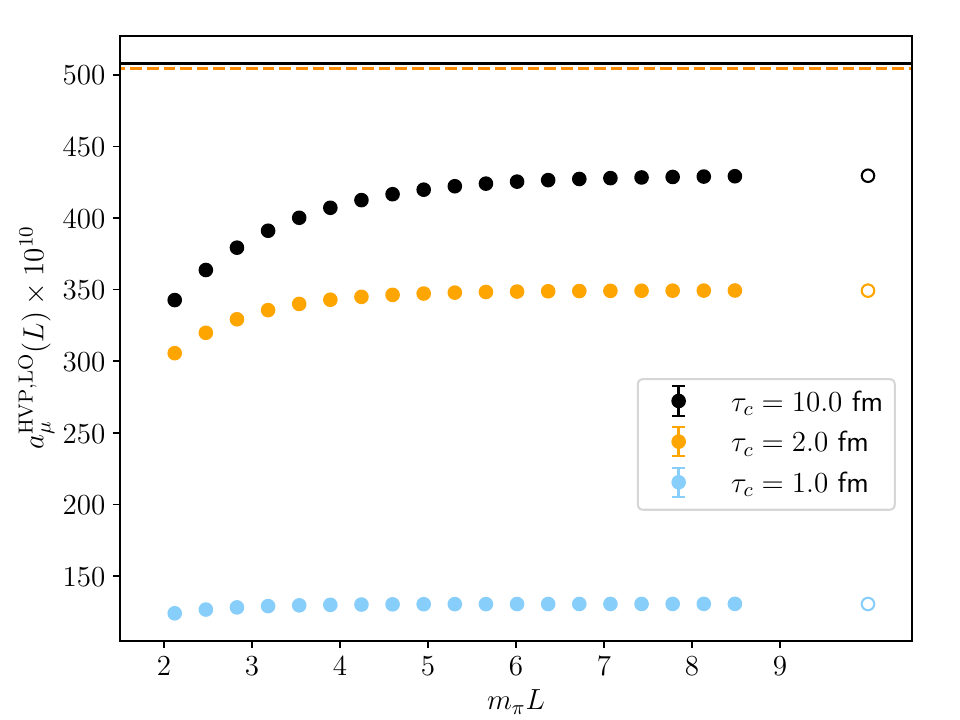}
   \caption{
   $\pi\pi$ contribution to $a_\mu^{\mathrm{HVP,LO}}$ for different volumes.
   The results for three different time integral upper limit are shown, {\it i.e.} $\tau_c=$ 1~fm, 2~fm and 10~fm.
   The infinite volume results are shown on the right by open symbols for each $\tau_c$.
   Two horizontal lines represent the estimates from dispersion analysis (solid line: DHMZ 19 \cite{Davier:2019can}, dashed line: KNT 19 \cite{Keshavarzi:2019abf}).
   }
   \label{fig:a_mu_v.s._m_piL}
\end{figure}

The results of the decomposition (\ref{eq:correlator_decomposed}) for $a_\mu^{\mathrm{HVP,LO}}$ are shown in Figure~\ref{fig:a_mu_v.s._m_piL}.
The results are shown for different values of the upper limit $\tau_c$ in the time-momentum integral (\ref{eq:time-momentum}): $\tau_c=$ 1~fm, 2~fm and 10~fm.
For the largest volumes, beyond $m_\pi L\gtrsim$ 6, we find that the finite-volume estimate is already close (within roughly $5\times 10^{-10}$) to the infinite volume limit (open circles).
    
In the same plot, the estimates from the dispersion relation using the experimental data of $e^+e^-\to\pi^+\pi^-$ are shown by horizontal lines; two groups, DHMZ 19 \cite{Davier:2019can} and KNT 19 \cite{Keshavarzi:2019abf}, are consistent with each other. 
The results of our analysis are significantly lower even with the largest $\tau_c=$ 10~fm, for which the integral has almost reached the value of $\tau_c\to\infty$.
In fact, the difference between $\tau_c=$ 5~fm and 10~fm is invisible at the scale of this plot.
The deficit compared to the dispersion method may be attributed to the GS model for the pion form factor.
As shown in Figure~\ref{fig:formfactor_exp_vs_GS}, the GS model significantly underestimates the experimental data near the $\rho$ resonance. However, this inconsistency would not largely affect our analysis of finite-volume effects since the agreement of the pion form factor is better in the lower-energy region, which is most responsible for finite-volume effects.

Figure~\ref{fig:a_mu_v.s._m_piL} clearly shows that the finite volume effect is significant for $m_\pi L\lesssim$ 6, which corresponds to $L\lesssim$ 8~fm, especially when the large time separation, $\gtrsim$ 1~fm, is included in the integral (\ref{eq:time-momentum}).
In the following section, we analyse the finite-volume effect $\Delta a_\mu^{\mathrm{HVP,LO}}$ in more details.





\section{Finite volume effects for \texorpdfstring{$a_\mu^{\mathrm{HVP,LO}}$}{amu}}
\label{sec:result}

In this section, we present our estimate of the finite-volume effects for $a_\mu^{\mathrm{HVP,LO}}$. The results for the total and sharply cut time windows are shown first, and those for the smeared windows follow.

\subsection{Total, and sharply cut time windows}
    
\begin{figure}[tb]
   \centering
   \includegraphics[width=0.9\linewidth]{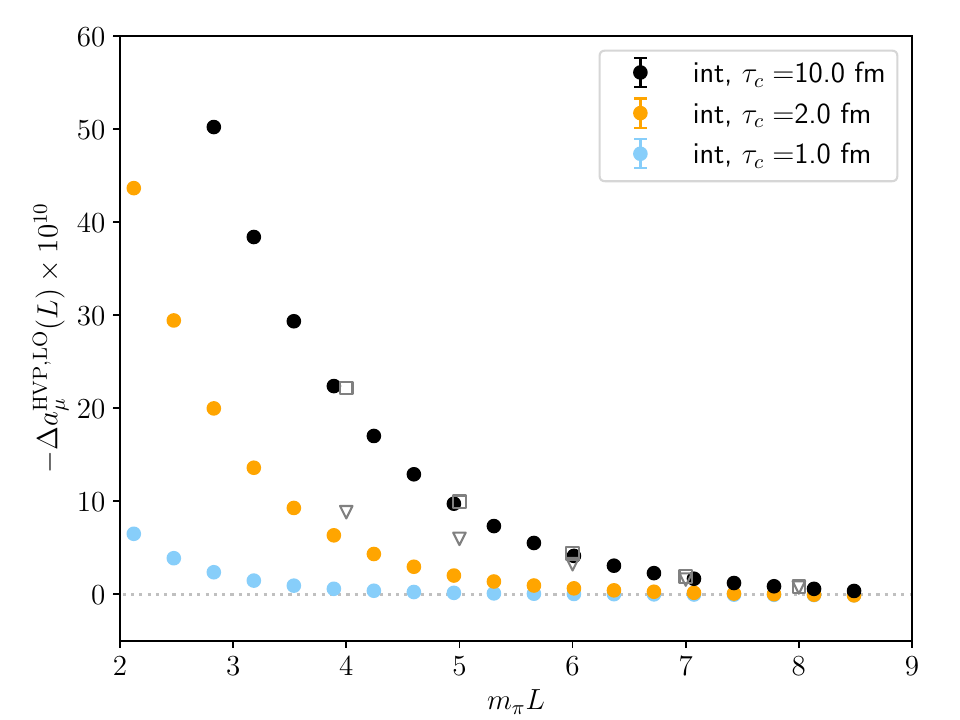}
   \caption{Finite volume effect for the muon anomalous magnetic moment $-\Delta a_\mu^{\mathrm{HVP,LO}}(L)$.
   Estimates are obtained with phenomenological $\pi\pi$ phase shift and time-like pion form factor described by the Gounaris-Sakurai model.
   Results are plotted for three values of upper limit $\tau_c$ in the Euclidean time integral (\ref{eq:time-momentum}): $\tau_c=$ 1~fm, 2~fm and 10~fm.
   The time integral is nearly saturated for $\tau_c=$ 10~fm.
   Estimates by Hansen and Patella \cite{Hansen:2019rbh,Hansen:2020whp} are also plotted (triangles: leading order, $e^{-m_\pi L}$ \cite{Hansen:2019rbh}, squares: including up to $e^{-3m_\pi L}$ \cite{Hansen:2020whp}).
   }
   \label{fig:delta_a_mu_v.s._m_piL}
\end{figure}

Figure~\ref{fig:delta_a_mu_v.s._m_piL} shows the finite-volume effect $-\Delta a_\mu^{\mathrm{HVP,LO}}(L)$ as a function of $m_\pi L$. The infinite-volume limit constructed with (\ref{eq:correlator_infiniteV}) is subtracted from the finite-volume results.

The finite-volume effect $-\Delta a_\mu^{\mathrm{HVP,LO}}(L)$ increases rapidly for $m_\pi L\simeq 6$ or smaller, especially when a region of large Euclidean time separation is included ($\tau_c$ = 2~fm and 10~fm). The size of $-\Delta a_\mu^{\mathrm{HVP,LO}}(L)$ is as large as $22\times 10^{-10}$ for $m_\pi L=4$, which is often considered a safe choice to suppress the systematic error due to the finite volume size. Since the target precision for $a_\mu^{\mathrm{HVP,LO}}$ is very high, {\it, i.e.} at the level of $5\times 10^{-10}$ or better, the size of the systematic effect as we find here has to be carefully estimated and subtracted.

The integral in the time-momentum representation (\ref{eq:time-momentum}) is saturated to an excellent precision at $\tau_c=$ 10~fm (black points in Figure~\ref{fig:delta_a_mu_v.s._m_piL}), and these data points can be considered an estimate of total $-\Delta a_\mu^{\mathrm{HVP,LO}}(L)$.

The estimates of Hansen and Patella obtained using effective field theory \cite{Hansen:2019rbh,Hansen:2020whp} are also plotted in Figure~\ref{fig:delta_a_mu_v.s._m_piL}.
They include the estimate of a wrap-around effect of the pion in one direction giving $e^{-m_\pi L}$ effects \cite{Hansen:2019rbh} (triangles), as well as those involving multiple wraps up to the term of $e^{-3m_\pi L}$ \cite{Hansen:2020whp} (squares).
(Contributions of order $e^{-\sqrt{2+\sqrt{3}}m_\pi L}$ are neglected, though.
It is therefore effectively summed up to $e^{-\sqrt{3} m_\pi L}$.)
These points can be compared with the full $-\Delta a_\mu^{\mathrm{HVP,LO}}(L)$, {\it i.e.} our estimate at $\tau_c=$ 10~fm.
Our results reproduce those of Hansen-Patella including $e^{-\sqrt{3} m_\pi L}$\cite{Hansen:2020whp}, implying that the both approaches are compatible for $m_\pi L\simeq$ 5--8. It also suggests that the wrap-around effect beyond those included is not significant for these volumes.

In the BMW calculation \cite{Borsanyi:2020mff}, the finite-volume effect is estimated in two steps.
Their reference lattice has a size $(L_{\mathrm{ref}}, T_{\mathrm{ref}})\simeq$ (6.3,9.4)~fm, which corresponds to $(m_\pi L_{\mathrm{ref}},m_\pi T_{\mathrm{ref}})\simeq$ (4.4,6.7).
From there, they estimate the difference from a bigger volume of $L_{\mathrm{big}}=T_{\mathrm{big}}\simeq$ 10.8~fm ($m_\pi L_{\mathrm{big}}=m_\pi T_{\mathrm{big}}\simeq$ 7.6), and then use chiral effective theory \cite{Hansen:2020whp} to estimate the difference from the infinite volume.
The first step, $L_{\mathrm{ref}}$ to $L_{\mathrm{big}}$, is estimated as $18(2)\times 10^{-10}$ mainly using actual simulation data on a coarse lattice, while the second is smaller than $1\times 10^{-10}$.
Our corresponding estimates are $13.7\times 10^{-10}$ and $1.0\times 10^{-10}$, respectively.
We find a reasonable agreement for both.
In the more recent publication \cite{Boccaletti:2024guq}, the estimate for $L_{\mathrm{ref}}\to L_{\mathrm{big}}$ from actual simulations is updated to $8.9(9)\times 10^{-10}$.
Their own estimate for $L_{\mathrm{ref}}\to\infty$ is also updated using a method similar to our work but with the pion form factor driven from Omnes formula as $9.1(2)\times 10^{-10}$, which is smaller than our analysis $14.8\times 10^{-10}$. The difference needs to be understood in detail.

In the recent calculation by the RBC/UKQCD collaboration \cite{Blum:2024drk} the largest volume lattice at the physical point has $m_\pi L=5.2$, for which we expect a correction of about $8.0\times 10^{-10}$. 

Another recent calculation by the Fermilab Lattice, HPQCD, and MILC Collaborations \cite{Bazavov:2024eou} employ lattices up to $L=$ 5.8~fm, which corresponds to $m_\pi L=4.0$. Our estimate of the finite-volume correction is about $20\times 10^{-10}$, but it may not be simply applied to their calculation using the staggered fermion, for which a dedicated formulation is required \cite{Aubin:2022hgm,FermilabLatticeHPQCD:2023jof}.

\begin{figure}[tb]
   \centering
   \includegraphics[width=0.9\linewidth]{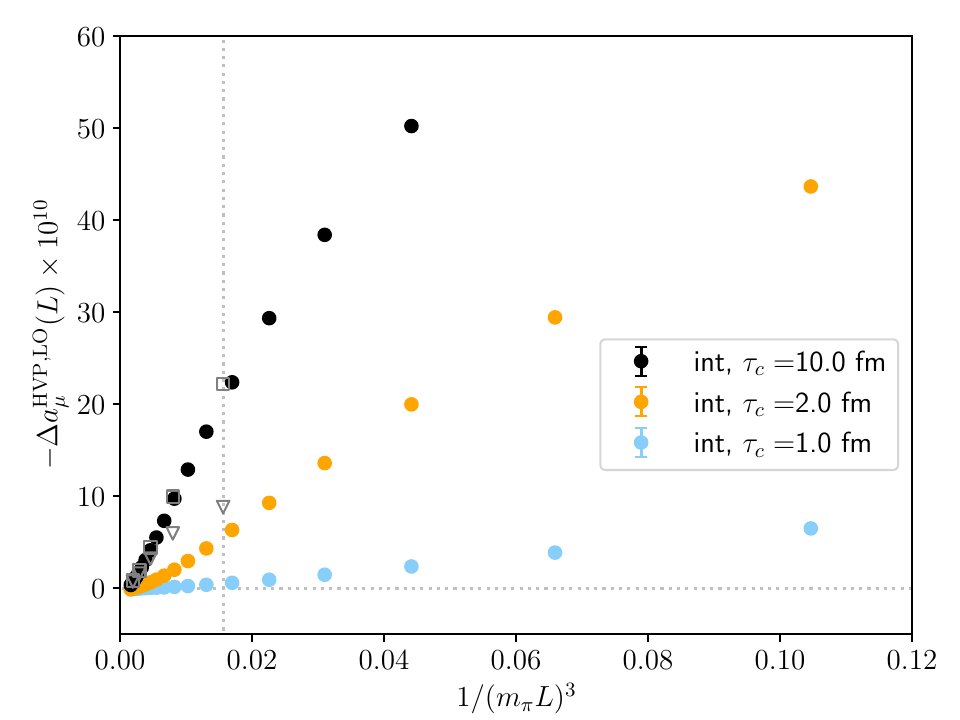}
   \caption{Finite volume effect for the muon anomalous magnetic moment $-\Delta a_\mu^{\mathrm{HVP,LO}}(L)$ plotted as a function of $1/(m_\pi L)^3$.
   The vertical dotted line shows $m_\pi L=4$.
   Other details are the same as in Figure~\ref{fig:delta_a_mu_v.s._m_piL}.
   }
   \label{fig:delta_a_mu_v.s._1/(m_piL)^3}
\end{figure}

The same results for $-\Delta a_\mu^{\mathrm{HVP,LO}}(L)$ are plotted as a function of $1/(m_\pi L)^3$ in Figure~\ref{fig:delta_a_mu_v.s._1/(m_piL)^3}.
Away from the large-volume regime, $1/(m_\pi L)^3\gtrsim$ 0.01, the finite-volume correction seems well described by a linear function of $1/(m_\pi L)^3$ for $\tau_c$ = 1.0, 2.0~fm. Including larger $\tau$, {\it i.e.} $\tau_c$ = 10.0~fm, the dependence is not that simple, but we can still see a component behaving as $1/(m_\pi L)^3$.
This reminds us of the possible $1/L^3$ dependence of $\Delta a_\mu^{\mathrm{HVP,LO}}(L)$ discussed in Sec.~\ref{sec:non-interacting_correlator}) for the non-interacting model.
Closer to the infinite volume limit, $1/(m_\pi L)^3\sim$ 0.01, they deviate from the linear dependence and approach zero more rapidly.

\begin{figure}[p]
    \centering
    \includegraphics[width=0.7\linewidth]{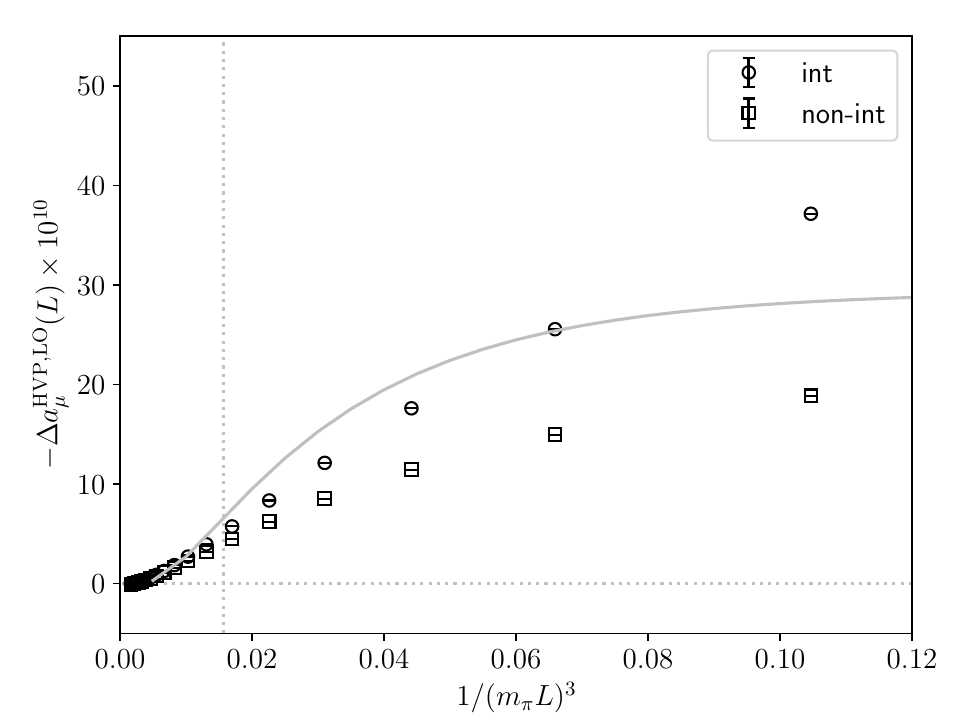}
    \includegraphics[width=0.7\linewidth]{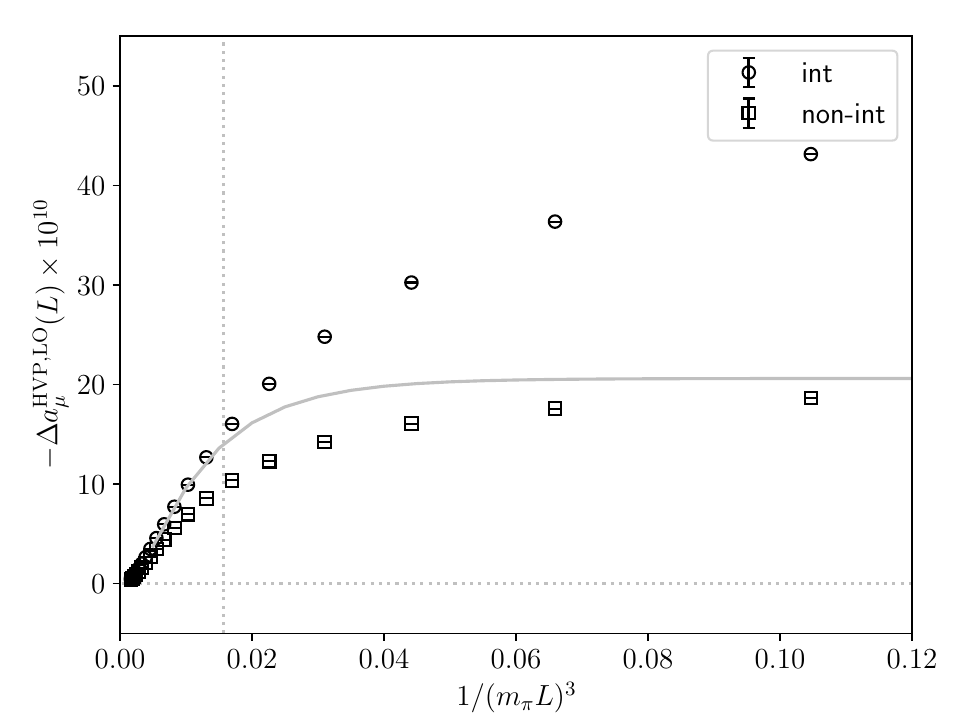}
    \caption{Finite volume effect for $a_\mu^{\mathrm{HVP,LO}}$ evaluated in the time range  [1,2]~fm (top panel) and in [2,10]~fm (bottom panel).
    Estimates with and without $\pi\pi$ interactions are shown (circles and squares, respectively).
    The error bar is from the extrapolation in the energy cutoff $E_{\pi\pi,\mathrm{cut}}$ (see Section~\ref{sec:sum_over_states}).
    The gray curve is obtained with the approximate formula for non-interacting $\Delta G(\tau,L)$, Eq.(\ref{eq:finite-V_correction_for_corr}), that is valid for $m_\pi\tau\gg 1$. 
    }
    \label{fig:delta_a_mu_v.s._1/(m_piL)^3_intvsnoint}
\end{figure}

To see this explicitly, we show a comparison between the finite-volume effect with and without the $\pi\pi$ interactions in Figure~\ref{fig:delta_a_mu_v.s._1/(m_piL)^3_intvsnoint}. The plots represent the contributions in the Euclidean time range [1,2]~fm (top panel) and [2,10]~fm (bottom panel). The non-interacting $\pi\pi$ contribution (squares) is evaluated numerically with inputs $\delta(k)=0$ and $F_\pi(s)=1$ and no additional approximation such as $m_\pi\tau\gg 1$ is involved. The approximate formula (\ref{eq:finite-V_correction_for_corr}) for $m_\pi\tau\gg 1$ (as shown in Figure~\ref{fig:damu_L3}) is also plotted by a gray curve. 
As mentioned associated with (\ref{eq:finite-V_correction_for_corr}), the non-interacting evaluation flattens for large $1/(m_\pi L)^3$ due to an accidental cancellation, while the interacting case shows a linear increase and the result turns out to be significantly larger compared to the non-interacting case.

Our result for $-\Delta a_\mu^{\mathrm{HVP,LO}}$ depends on the details of the $\pi\pi$ interaction, {\it i.e.} the phase shift and time-like pion form factor. In order to see the size of its dependence, we repeat the analysis with slightly modified parameters in the Gounari-Sakurai model. That is, we shift the $\rho$ meson mass parameter $m_\rho$ by $-20/+20$~MeV, and check how much it affects the resulting $-\Delta a_\mu^{\mathrm{HVP,LO}}$. The results are $+2.9/-2.6$, $+0.6/-0.6$, $+0.14/-0.11$, and $+0.05/-0.06$ for $L=$ 3, 5, 7 and 10~fm, respectively in unit of $10^{-10}$. Compared to the finite-volume effect itself, the shift gives a minor effect.

\subsection{Smeared time windows}
\label{sec:smeared_time_windows}

For convenience, we also provide the results for the smeared time windows proposed by \cite{RBC:2018dos} and commonly adopted by many collaborations.
Using the smeared Heaviside function
\begin{equation}
   \Theta(\tau,\tau';\Delta)\equiv \frac{1}{2} \left[ 1+\tanh\frac{\tau-\tau'}{\Delta} \right],
\end{equation}
the leading-order HVP contribution $a_\mu^\mathrm{HVP,LO}$ can be decomposed into short-distance (SD), window (W) and long-distance (LD) pieces as $a_\mu^{\mathrm{HVP,LO}}=a_\mu^{\mathrm{SD}}+a_\mu^{\mathrm{W}}+a_\mu^{\mathrm{LD}}$.
Their time-momentum representations of (\ref{eq:time-momentum}) are given as
\begin{widetext}
\begin{align}
   a_\mu^\mathrm{SD}
   &= 4\alpha^2 m_\mu \int_0^{\tau_c} d\tau\,
   [1-\Theta(\tau,\tau_0;\Delta)]\,
   \tau^3 G(\tau) \tilde{K}_E(\tau),
   \nonumber\\
   a_\mu^\mathrm{W}
   &= 4\alpha^2 m_\mu \int_0^{\tau_c} d\tau\,
   [\Theta(\tau,\tau_0;\Delta)-\Theta(\tau,\tau_1;\Delta)]\,
   \tau^3 G(\tau) \tilde{K}_E(\tau),
   \nonumber\\
   a_\mu^\mathrm{LD}
   &= 4\alpha^2 m_\mu \int_0^{\tau_c} d\tau\,
   \Theta(\tau,\tau_1;\Delta)\,
   \tau^3 G(\tau) \tilde{K}_E(\tau).
   \label{eq:time-momentum_W}
\end{align}
\end{widetext}
The separation points are set as $\tau_0=$ 0.4~fm and $\tau_1=$ 1.0~fm, and the width $\Delta$ of the smearing is set to 0.15~fm.
We take $\tau_c$ = 10~fm, which nearly saturate the integral up to $\infty$.

\begin{figure}[tb]
\centering
\includegraphics[width=0.9\linewidth]{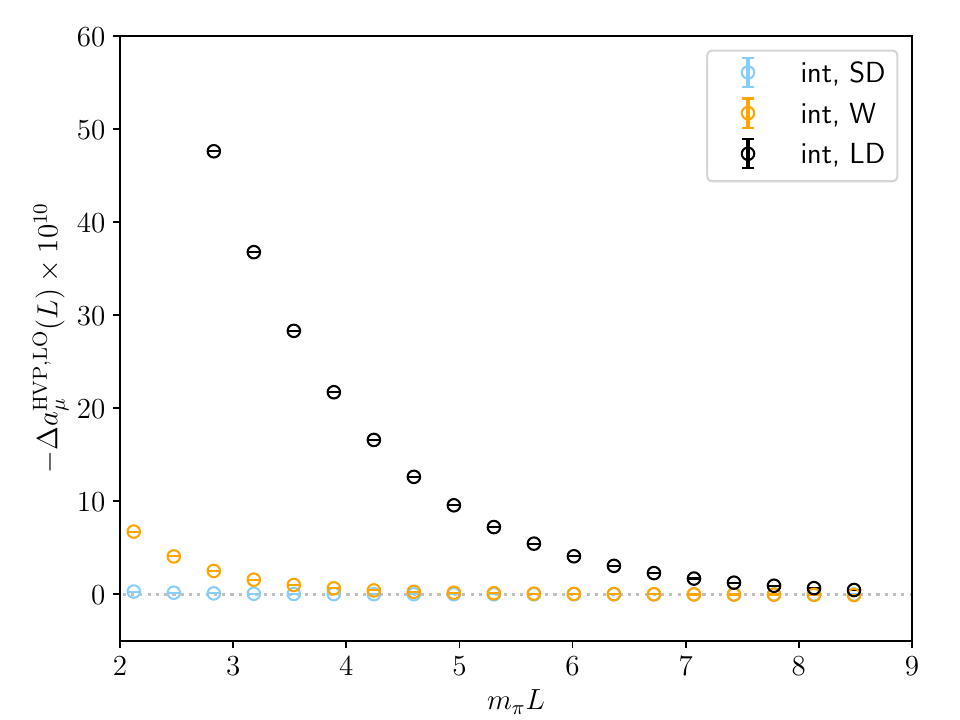}
\caption{Finite-volume effect for $a_\mu^{\mathrm{HVP,LO}}$ evaluated for different Euclidean time windows, plotted as a function of $m_\pi L$.
The error bar is from the extrapolation in the energy cutoff $E_{\pi\pi,\mathrm{cut}}$ (see Section~\ref{sec:sum_over_states}).
}
\label{fig:delta_a_mu_v.s._m_piL_windows}
\end{figure}

\begin{figure}[tb]
\centering
\includegraphics[width=0.9\linewidth]{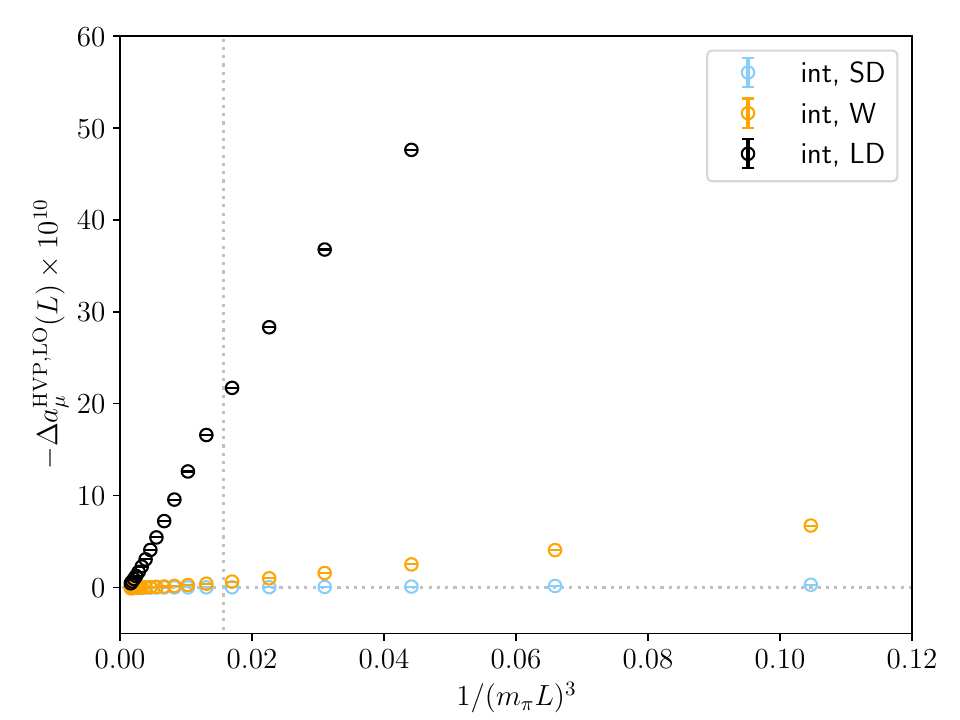}
\caption{Finite-volume effect for $a_\mu^{\mathrm{HVP,LO}}$ 
evaluated for different Euclidean time windows, plotted as a function of $1/(m_\pi L)^3$.
The error bar is from the extrapolation in the energy cutoff $E_{\pi\pi,\mathrm{cut}}$ (see Section~\ref{sec:sum_over_states}).
}
\label{fig:delta_a_mu_v.s._1/(m_piL)^3_windows}
\end{figure}

Similarly to Figures~\ref{fig:delta_a_mu_v.s._m_piL} and \ref{fig:delta_a_mu_v.s._1/(m_piL)^3}, we plot the estimates of the finite-volume effect for each time window in Figure~\ref{fig:delta_a_mu_v.s._m_piL_windows} (versus $m_\pi L$) and in Figure~\ref{fig:delta_a_mu_v.s._1/(m_piL)^3_windows} (versus $1/(m_\pi L)^3$).
Different windows are distinguished by different colors.
The finite-volume effect is most significant for the long-distance (LD) window, and its dependence on $m_\pi L$ is very similar to those of sharp-cut windows.

The numerical values are listed in Table~\ref{tab:delta_a_mu_windows}.
The errors are from the extrapolation in the energy cut-off $E_{\pi\pi,\mathrm{cut}}$ (see Section~\ref{sec:sum_over_states}).
Other sources of errors, such as those from the input parameters as well as from the ignored inelastic states, are not shown.

\begin{table}[tb]
    \centering
    \begin{tabular}{D{.}{.}{8}D{.}{.}{8}D{.}{.}{8}D{.}{.}{8}D{.}{.}{8}}
    \hline
    {\text{$L$ [fm]}} & {\text{SD}} & {\text{W}} & {\text{LD}} & {\text{total}}\\
    \hline
    3.0 &
    0.293 & 6.737 & 79.756 & 86.787\\
    3.5 &
    0.166 & 4.066 & 61.558 & 65.790\\
    4.0 &
    0.097 & 2.508 & 47.611 & 50.216\\
    4.5 &
    0.058 & 1.571 & 36.769 & 38.397\\
    5.0 &
    0.035 & 0.997 & 28.311 & 29.343\\
    5.5 &
    0.021 & 0.639 & 21.715 & 22.375\\
    6.0 &
    0.013 & 0.413 & 16.588 & 17.013\\
    6.5 &
    0.008 & 0.267 & 12.621 & 12.896(1)\\
    7.0 &
    0.004 & 0.170 & 9.562 & 9.736(1)\\
    7.5 &
    0.001 & 0.109(1) & 7.221 & 7.331(2)\\
    8.0 &
    0.000(1) & 0.071(2) & 5.443 & 5.513(3)\\
    8.5 &
    -0.002(1) & 0.042(3) & 4.085 & 4.125(5)\\
    9.0 &
    -0.003(1) & 0.020(4) & 3.055 & 3.072(7)\\
    9.5 &
    -0.005(2) & 0.001(6) & 2.276(1) & 2.272(10)\\
    10.0 &
    -0.007(2) & -0.014(9) & 1.690(2) & 1.669(14)\\
    10.5 &
    -0.009(3) & -0.028(12) & 1.247(4) & 1.210(19)\\
    11.0 &
    -0.011(3) & -0.043(15) & 0.916(6) & 0.863(25)\\
    11.5 &
    -0.013(4) & -0.062(20) & 0.659(8) & 0.584(33)\\
    12.0 &
    -0.017(5) & -0.086(24) & 0.458(11) & 0.355(41)\\
    \hline
    \text{$\infty$} &
    12.31 & 120.3 & 296.9 & 429.5\\
    \hline
    \end{tabular}
    \caption{$-\Delta a_\mu^{\mathrm{HVP,LO}} \times 10^{10}$ for three Euclidean time windows, short-distance (SD), intermediate (W) and long-distance (LD), as well as a sum over them corresponding to the total contribution.
    The errors are from the extrapolation in the energy cutoff $E_{\pi\pi,\mathrm{cut}}$ (see Section~\ref{sec:sum_over_states}).
    If the error is not written, it is smaller than the last digit shown.
    The last line is $a_\mu^{\mathrm{HVP,LO}} \times 10^{10}$ in the infinite volume.
    }
    \label{tab:delta_a_mu_windows}
\end{table}

\section{Conclusion}
\label{sec:conclusion}

In this paper, the finite-volume effect for the contribution of leading-order hadronic vacuum polarization to the muon $g-2$, $a_\mu^{\mathrm{HVP,LO}}$, is studied in detail using phenomenological inputs.
Unlike previous works \cite{Aubin:2015rzx,Hansen:2019rbh,Hansen:2020whp}, we explicitly construct the Euclidean correlator in a box assuming that only two-pion states contribute without any truncation of the pion's wrapping-around effect.
The phenomenological inputs are the $\pi\pi$ phase shift and the time-like pion form factor, both of which can be consistently parametrized using the Gounaris-Sakurai model.

The $\pi\pi$ states in this analysis satisfy L\"uscher's quantization condition, and their energy spectrum automatically reflects the wrapping-around effect to all orders, in contrast to previous studies that include only the leading terms.
The Euclidean correlator $G(\tau,L)$ at a finite volume $L$ thus constructed shows a different volume dependence for different Euclidean time separations $\tau$.
Asymptotically, the infinite volume limit is approached as $\exp(-m_\pi L^2/4\tau)$, while an approximate power scaling $1/L^3$ is expected in the intermediate volume region in general.

The hadronic vacuum polarization contribution to $a_\mu^{\mathrm{HVP,LO}}$ is obtained by a weighted integral over Euclidean time dominated in the region around $\tau\sim$ 1~fm.
The overall $L$ dependence is then not simply described by a single functional form.
It turns out that the finite-volume effect $\Delta a_\mu^{\mathrm{HVP,LO}}$ shows an approximately linear dependence on $1/(m_\pi L)^3$ in the region $1/(m_\pi L)^3 > 0.01$, and rapidly decreases in the larger volume region.
Their size is consistent with the estimate reported by Hansen and Patella \cite{Hansen:2020whp}.

Our analysis fills the gap between the exponential finite-volume scaling based on {\it e.g.} chiral effective theory and the power-law dependence expected from L\"uscher's quantization condition. Such a complication may arise for other quantities, for which the lowest-lying part of the spectrum is given by two-body states, although it became most critical for $a_\mu^{\mathrm{HVP,LO}}$ because of its high required precision.

Similar analysis can be done for heavier pions employed in many lattice calculations, provided that the phase shift and pion form factor are available. In other words, these quantities will provide a means of estimating finite-volume effects in the lattice calculations of $a_\mu^{\mathrm{HVP,LO}}$. When the lattice data are available on multiple volumes, this analysis would provide a highly non-trivial cross-check, as it is based on rather solid ground at least for the finite-volume effects.

Finally, we note that our analysis is not complete as it could not take account of the inelastic states, such as four pions, and the realistic estimate can only be obtained with the full scale simulations such as those of \cite{Borsanyi:2020mff,Boccaletti:2024guq,Blum:2024drk}.
However, the most important contribution from the $\pi\pi$ states would be well captured by our analysis.

\begin{acknowledgments}
We thank Kohtaroh Miura for numerous discussions.
We are grateful to Haoyu Wang for offering a detailed numerical check for some of our results.
SH is partly supported by MEXT as ``Program for Promoting Researches on the Supercomputer Fugaku'' (JPMXP1020200105) and by JSPS KAKENHI, Grant-Number 22H00138.
\end{acknowledgments}


\nocite{*}

\bibliography{biblio}

\end{document}